\documentclass[%
 reprint,
 amsmath,amssymb,
 aps,
pra,
]{revtex4-2}

\usepackage{graphicx}
\usepackage{dcolumn}
\usepackage{bm}

\usepackage{diagbox}
\usepackage{multirow}
\usepackage{hyperref}

\usepackage{xcolor}

\usepackage{CJKutf8}

\begin{document}

\begin{CJK}{UTF8}{gbsn}

\title{Topological phase transition induced by modulating unit cells in photonic Lieb lattice}
\author{{Zhi-Kang Xiong}\textsuperscript{1}}
\author{{Y. Liu (刘泱杰)}\textsuperscript{1}}
\email[Corresponding author: ]{yangjie@hubu.edu.cn}
\author{{Xiying Fan}\textsuperscript{1}}
\author{{Bin Zhou}\textsuperscript{1, 2}}
\email[Corresponding author: ]{binzhou@hubu.edu.cn}
\affiliation{
$^{1}$ Department of Physics, School of Physics, Hubei University, Wuhan 430062, P. R. China\\
$^{2}$ Wuhan Institute of Quantum Technology, Wuhan 430206, P. R. China}%

\date{v71, submitted to New J. Phys. 17th Aug., major revision requested 18th Dec. '25, resubmitted 13th Jan. '26. }

\begin{abstract}

Topological photonics was embarked from realizing the first-order chiral state in gyromagnetic media, but its higher-order states were mostly studied in dielectric lattice instead. In this paper we theoretically unveil a hierarchy of topological phases under broken time-reversal symmetry, which include the first-order Chern, and the second-order dipole, quadrupole phases. Concretely, by relaxing a certain spatial symmetry of unit cell, versatile topological phases including both edge and corner states can be established to transit around, with bandgap closures marking the phase boundaries. 
Our results on gyromagnetic photonic crystals may broaden the scope of sublattice engineering design for topological phase manipulation, potentially enabling multifunctional disorder-resistant waveguides and integrated photonic circuits for information communication.


\end{abstract}

\maketitle

\end{CJK}


\section{Introduction}

Topological phase, recently explored in condensed matter physics, features robust low-dimensional edge states undisturbed by structural disorder \cite{hasan2010,qi2011,klitzing1980,kane2005,shen2012}. The topological phase also inspired topological photonics as a new arena to play in~\cite{lu2014t,ozawa2019, khanikaev2013,haldane2008,wang2008}. The bulk-edge correspondence~\cite{hasan2010} dictates that photonic crystals (PhCs) with miscellaneous Chern numbers for their bands~\cite{wang2008,wang2009, ma2016, he2019, yu2022topological, wu2015, ren2023topological} host protected edge channels in optical waveguides \cite{yu2018,wu2019}. These systems then offer a novel platform to emulate the transition governed by such \emph{first-order} topological invariants~\cite{shen2012}.
Among them, a simple way to exert topological phase transitions in reciprocal space is to modulate the unit cell accordingly in real space, so that the intercell hopping strength dominates over the intracell one, as seen in the one-dimensional SSH model~\cite{su1979solitons, heeger1988solitons}. Alternatively, rotating the unit cell to vary the sublattice distance can achieve a similar effect in PhC~\cite{zhou2021topological}. Such experimental proposals for topological PhCs can be exploited in designing laser devices~\cite{han2019, sun2021, morita2021} to achieve stable and efficient output. And they even facilitate a nonlinear imaging method with the third-harmonic generation~\cite{kruk2019nonlinear, smirnova2019, lan2020nonlinear}, whose signals for bulk and edge states are measurable at high contrast for a wide frequency band~\cite{smirnova2019}. Furthermore, topological PhCs also contribute to the field of quantum optics to generate and control quantum states~\cite{barik2020chiral, chen2021topologically}. 

As an extension to the first-order Chern phase, higher-order topological insulators are noted with further progress, in which the dimension $d$ of the topological boundary state is at least two orders less than that ($n$) of the bulk, i.e., $n-d\geqslant2$~\cite{song2017d, benalcazar2017electric}. Firstly in 2017, Benalcazar, Bernevig, and Hughes (BBH) proposed a higher-order topological concept to quantize electric multipole moments in crystalline insulators, such as dipole, quadrupole, and octuple moments~\cite{benalcazar2017electric,benalcazar2017quantized}. Their prediction was soon emulated in photonic crystals, which simultaneously support both one-dimensional edge states and zero-dimensional corner states~\cite{xie2018,el2019,xie2019,he2020}. In general, the topological corner states can live in systems with non-trivial bulk dipole moment, as well as with non-trivial quadrupole moment~\cite{benalcazar2017quantized,benalcazar2017electric, lan2023}. To date, second-order corner states enable diverse application scenarios, including nonlinear photonics~\cite{kirsch2021,smirnova2020}, topological high-Q resonances~\cite{zangeneh2019,kang2022}, nanolasers~\cite{zhang2020low}, and topological photonic crystal fibers \cite{lin2020,zhang2021}. To induce such corner states, one can modulate the unit cells in the concerned lattice, e.g. in a square lattice with ${\rm C}_4$ symmetry~\cite{xie2021higher, chen2022, vaidya2023, lan2024}, similar to rotating unit cells in a dielectric PhC~\cite{zhou2021topological, ZhouR2022valley_vortices} mentioned above. One may also refer to a review on higher-order topological phases in crystalline and
non-crystalline systems~\cite{YangY2024Higher-order}.

Now we ask further whether breaking time-reversal symmetry and ${\rm C}_4$ symmetry of a primitive Lieb lattice~\cite{nictua2013, guzman2014, mukherjee2015} can host higher-order topological states similar to these in the square lattice.
Previous work demonstrated a quadrupole phase in dielectric PhC~\cite{lan2023}. Instead here in our paper, a gyromagnetic PhC~\cite{belotelov2005,wang2009} is adopted to open more band gaps in a 2D Lieb lattice with broken time-reversal symmetry. This approach yields both the first-order and second-order topological states, resulting in a larger variety of phases than its dielectric counterpart. In the Lieb lattice, a unit cell carries three unbalanced sublattices~\cite{leykam2018}, providing a unique pattern of periodic lattice enabling an extra degree of freedom to twist around. We find out that while Chern and quadrupole phases already emerge in a primitive Lieb PhC with uniform radii, a deformation with varied radii or shifted distances for sublattices induces a hierarchy of Chern, dipole, and quadrupole phases. Our results hence pin down multiple topological invariants in the deformed Lieb PhC, which shall shed new light on higher-order phases which are readily applicable to design photonic structures for information communication. We note that more topological gaps result from \emph{broken time-reversal symmetry} than from its time-reversal symmetric counterpart in our lattices. 

\section{PhC and Wilson-loop calculation}
In this section, we will introduce the primitive Lieb lattice and its deformations of our concern in this paper in Subsec.~\ref{Primitive}, and the calculation method of the relevant topological invariants in Subsec.~\ref{Calculation}. 

\subsection{The Lieb lattices: primitive and deformed \label{Primitive}}
\begin{figure}[hbtp!]
\includegraphics[width=0.48\textwidth]{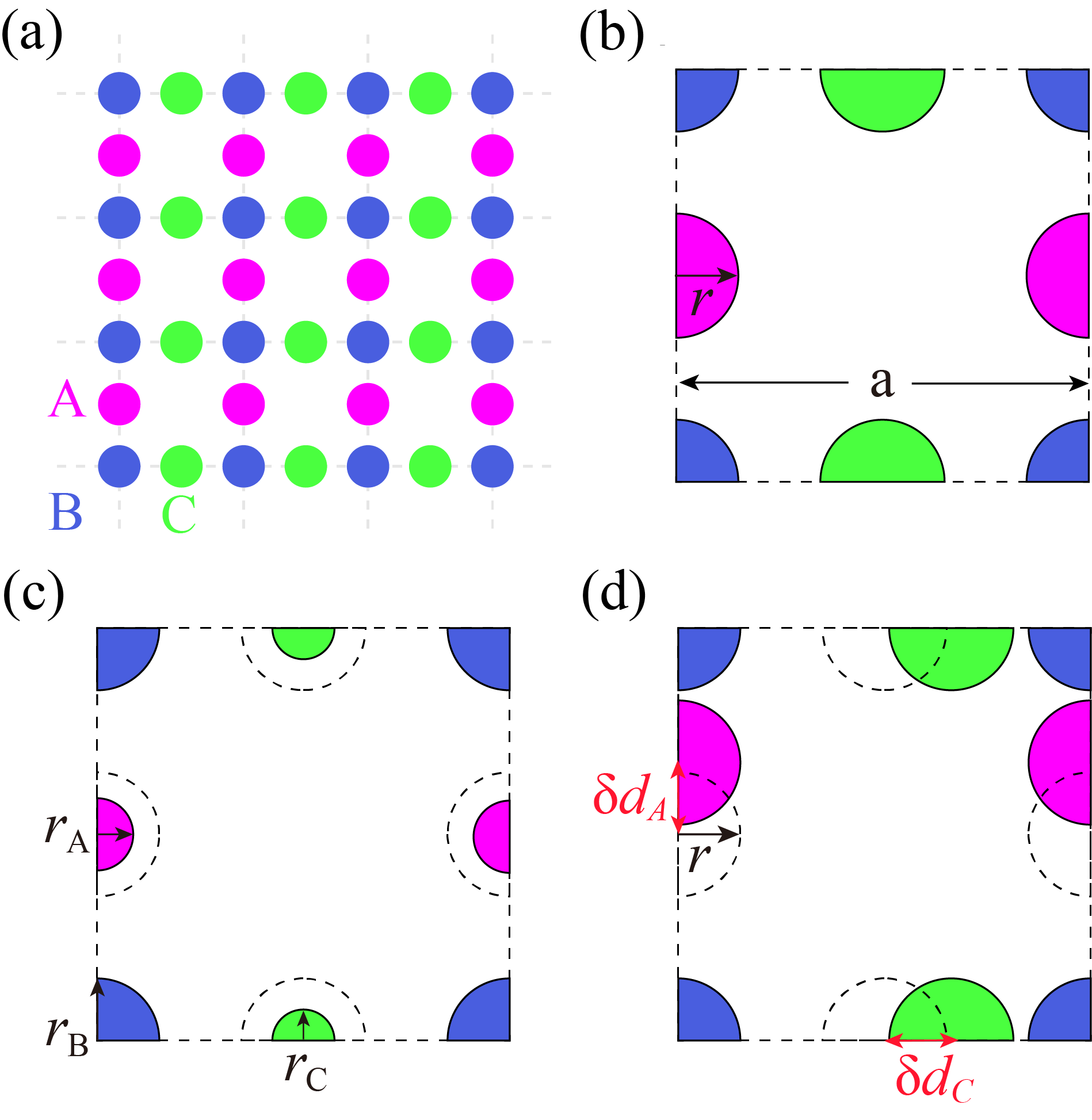}
\caption{\label{fig:fig1}(a) Schematic for the primitive Lieb lattice consisting of YIG rods in air. The labels A, B and C represent three disks as sublattices. (b) Unit cell for the primitive lattice, with uniform radius $r$ of the gyromagnetic sublattices. (c) Deformed unit cell with non-uniform radii $r_A, r_B, r_C$ of three sublattices. (d) Deformed unit cell with shifted sublattices A and C, for which the shifted distances are represented by $\delta d_A$ and $\delta d_C$ respectively.   
}
\end{figure}

The Lieb lattice in primitive form throughout this paper is shown in Fig.~\ref{fig:fig1}(a), where A, B, and C respectively represent three distinct sublattices~\cite{guzman2014}. All the sublattices represent gyromagnetic cylinders made of yttrium-iron-garnet (YIG) rods \cite{wang2009}. The permittivity of such a YIG rod is $\epsilon=15\epsilon_0$, and the permeability tensor is
\begin{eqnarray}
\bar{\bar{\mu}}= \left[\begin{array}{ccc}
\mu&i\kappa& 0\\
-i\kappa&\mu&0\\
0&0&\mu_0
\end{array}\right], \label{0}
\end{eqnarray}
where $\mu=14\mu_0$ and $\kappa=12.4$, $\epsilon_0$ and $\mu_0$ are the vacuum permittivity and permeability, respectively. The unit cell for primitive Lieb lattice is shown in Fig.~\ref{fig:fig1}(b), where $a$ is the lattice constant, and $r$ is the radius for the gyromagnetic cylinders. It is the magneto-optical effect from Eq.~\eqref{0} that breaks the time reversal symmetry of the system and results in a non-zero Chern number in its band~\cite{haldane2008,wang2008, wang2009}. 
One can also interpret such a Lieb PhC in a Hamiltonian form with unbalanced coupling strengths between the three neighbouring sublattices A, B, and C~\cite{chen2017spin, jiang2019topological}. To induce more higher-order phases, we exploit its two spatial degrees of freedom to deform unit cells of the primitive Lieb PhC:  (1) varying the radii of sublattices as shown in Fig.~\ref{fig:fig1}(c); (2) shifting the positions of sublattices as shown in Fig.~\ref{fig:fig1}(d). We then find out various topological phases, such as the first-order Chern phases and second-order quadrupole phases for three cases: uniform radii of sublattices, varied radii of sublattice A (or C), and shifted positions of sublattices in the unit cells, which are presented in Secs.~\ref{uniform}, ~\ref{inhomogeneous}, and~\ref{shift} respectively. 

\subsection{Calculation method for Wilson loop and nested Wilson loop~\label{Calculation}}
In this subsection, we briefly introduce the methods of Wilson loop and nested Wilson loop developed by Benalcazar~\emph{et al.}~\cite{benalcazar2017electric} to characterize the topological phases of our Lieb lattices. We shall outline the main procedures to compute the Chern number, dipole moments, and the quadrupole moments as follows (also see Append.~\ref{Method} for further detail). 

We begin by numerically solving the band structure and eigenmodes under periodic boundary conditions (PBCs). For transverse magnetic (TM) modes, the Bloch function for the $n$-th band is defined by the field profile of $E_z^n(k)$, i.e., $u^n_k(\textbf{r})\equiv e^{-i\textbf{k}\cdot\textbf{r}}E_{z,k}^n(\textbf{}r)$~\cite{blanco2020tutorial}.

For band $n$, its Chern number $C_n$ is calculated in the reciprocal space over the first Brillouin zone (BZ). Here we discretize the first BZ into a series of small square units, calculate the Berry curvature for each unit, and then integrate over the entire first BZ to obtain the Chern number of the corresponding energy band. Then we use a Wilson-loop-based approach to calculate the dipole and quadrupole moments, with the photonic band structure and eigenmodes obtained from full-wave numerics. First, the eigenfrequencies and electric field profiles $E_z(\textbf{r})$ of the TM modes are solved for over a discretized mesh of the first BZ. Second, all eigenmodes are normalized using the dielectric-weighted inner product: $
\langle u_m(\textbf{k})|\epsilon(\textbf{r})|u_n(\textbf{k})\rangle=\delta_{m,n}$. The first BZ is discretized into a uniform grid of $40\times40$ in most cases, the spacing in reciprocal space is $\Delta k_x=\Delta k_y=2\pi/aN$ (see Append.~\ref{Method} for further detail).

The dipole moment originates from the non-trivial Berry phase of the energy band or the polarization of the Wannier center. For trivial polarization, the Wannier centers are evenly distributed in the unit cell without overall shift of polarization. However, non-trivial polarization are instead shifted by a certain distance, causing mode localization. Concretely in the Lieb lattice, the quantized dipole moment $P_i(i=x, y)=0$ or $0.5$ represents trivial and non-trivial polarization respectively with the reflection symmetry $M_i$ and $C_4$ symmetry (see Append.~\ref{symmetry}). Each of the two components $P_x$ and $P_y$ of the polarization independently describes the edge state.

Moreover, the bulk quadrupole moment can also characterize the edge and corner states of the Lieb lattice~\cite{lan2023}. For the quadrupole moment, it can be defined only when the total dipole polarization $P_i$ vanishes~\cite{benalcazar2017electric}. The quadrupole moment is defined by the polarization of the Wannier sector, which is related to the nested Berry phase and can be understood as the second-order polarization of the Wannier center. Non-trivial bulk quadrupole moments $q_{xy}=0.5$ in this Lieb lattice are already pinned down in previous literature~\cite{lan2023}. 

\section{Topological phases in primitive Lieb lattice: uniform radii for three sublattices} \label{uniform}

\begin{figure*}[hbtp!]
\includegraphics[width=0.96\textwidth]{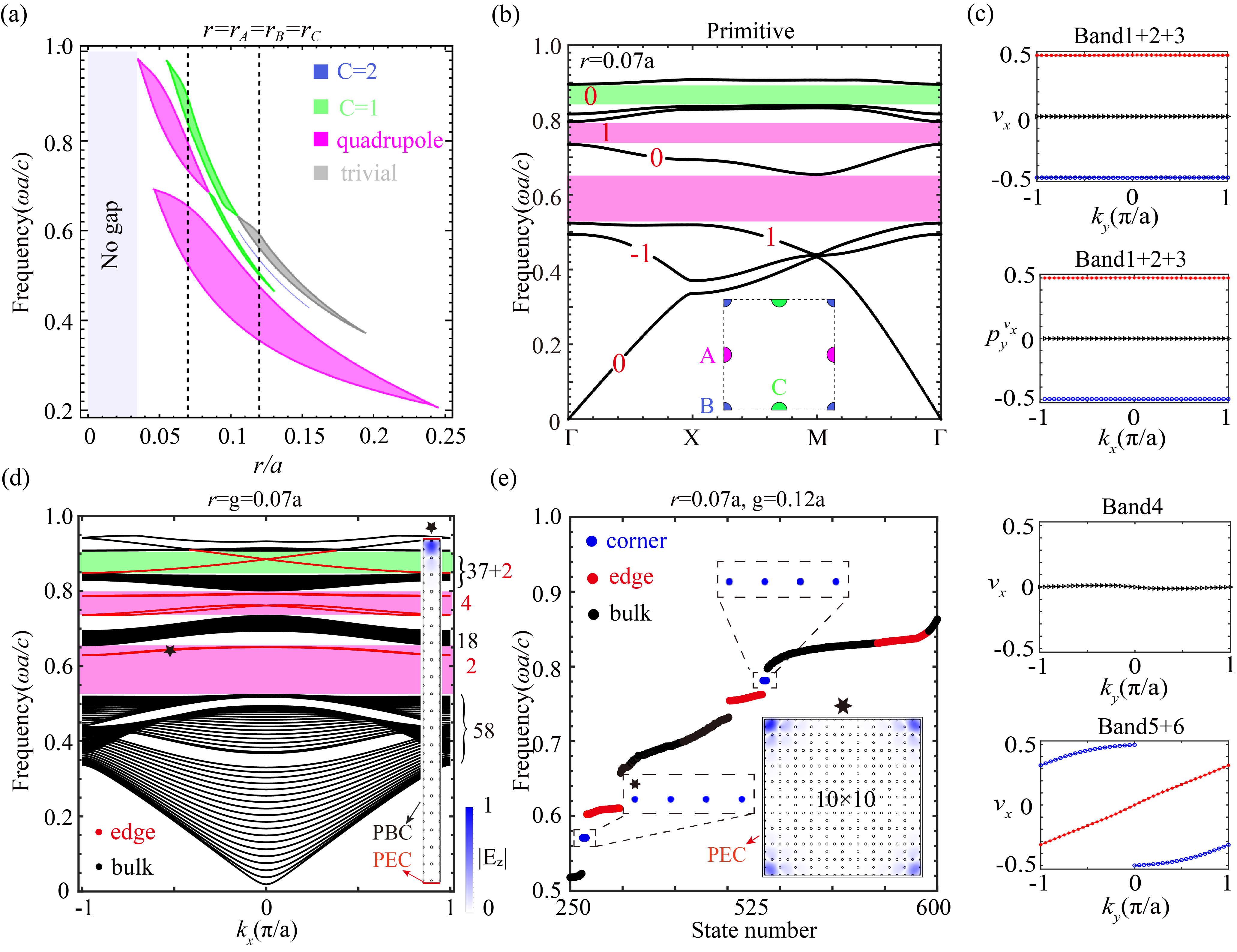}
\caption{\label{fig:fig2}(a) The band gap diagram of the primitive lattice with various topological phases as the radius $r$ of three sublattices increases. The coloration for each topological phase is indicated in its legend. (b) Band structure of the gyromagnetic Lieb PhC with radius $r=0.07a$, with the Chern number of each band marked. The inset represents the unit cell structure of the primitive lattice, and the coloration for phase follows the legend of panel (a). 
(c) The Wannier center distributions of each band in four panels as follows. Top panel: the Wannier center $v_x$ of the first band gap with trivial dipole polarization $P_x=0$, upper panel: the polarization of the Wannier center $p_y^{v_x}$ of the first band gap with non-trivial quadrupole moment; lower panel: the Wannier center $v_x$ of the fourth band with trivial polarization, bottom panel: $v_x$ of the fifth and sixth bands with non-trivial Chern phase with $C=1$. 
(d) Projected band diagram of a super cell consisting of $1\times 20$ unit cells, with $r=g=0.07a$. The red curves represent the edge states in gaps. The inset shows the electric field profile $E_z$ for the point marked by the pentagram. (e) Eigenstate diagram of a super cell consisting of $10 \times10$ unit cells with $r=0.07a, g=0.12a$, bounded by four OBCs along with an air gap beneath. The inset shows the electric field $E_z$ for the corner state marked by the black hexagram. 
}
\end{figure*}

In this section, we consider the primitive Lieb lattice, i.e., with uniform radii of three sublattices, $r_A=r_B=r_C=r$ as shown in Fig.~\ref{fig:fig1}(b). As the uniform radius $r$ for three sublattices changes, rich topological phases occur such as Chern phase ($C=1, 2$) and quadrupole phase ($q_{xy}=0.5$), as charted in Fig.~\ref{fig:fig2}(a). We shall present the phase transition for the primitive Lieb lattice as below.

The coloured areas in Fig.~\ref{fig:fig2}(a) map the band gaps versus the sublattice radius $r$, showing the topological invariants marked for the gaps. Band gaps appear in our Lieb lattice only when the radius reaches $0.035a$. The band gaps are speculated to occur only when the lattice scatters light strongly enough, which requires $r>0.035a$ but no more than $r=0.25a$ limited by touching between pillars in the unit cell. Of them, the first and second gaps in magenta both close when the radius $r$ gets large enough to $0.25a$ and $0.085a$ respectively, similar to the dielectric counterpart~(also two gaps in Fig.~1 in~\cite{lan2023} though the second gap is not presented there). And the third gap between $0.63\sim 0.98 \omega a/c$ with Chern number $C=1$ opens uniquely due to the gyromagnetic PhC, which would not have existed in its dielectric counterpart~\cite{lan2023}. All across Fig.~\ref{fig:fig2}(a), corner states prevail for the radius range in $0.035a<r<0.25a$ and Chern phase with $C=1$ comes in for the range of $0.055a<r<0.13a$, Chern phase $C=2$ in $0.1a<r<0.15a$.

We select two cases of $r=0.07a$ and $r=0.12a$ for concreteness, which are marked with black dashed lines in Fig.~\ref{fig:fig2}(a). In the case of $r=0.07a$, the band structure for a primitive lattice [see its inset, the same as Fig.~\ref{fig:fig1}(b)] gives three gaps in Fig.~\ref{fig:fig2}(b), which are marked in colour with corresponding phases in Fig.~\ref{fig:fig2}(a). 
For the first three bands below the first gap in Fig.~\ref{fig:fig2} (b), their Wannier center $v_x(k_y)$ adds up to $P_x=0, P_y=0$~\footnote{And the dipole polarization $P_y$ is the same as $P_x$ due to the $C_4$ symmetry: $P_x=P_y$ [See Eq.~\eqref{C4}, Append.~\ref{symmetry}]} from Eq.~\eqref{px}, which shows trivial dipole polarization in top left of Fig.~\ref{fig:fig2}(c). Among them, the first and third bands exhibit $v_x=\pm0.5$ while the second band is trivial with $v_x=0$. Thus the polarization of the Wannier center $v_x$ characterizes a non-trivial quadrupole moment $q_{xy}=0.5$ from Eqs.~\eqref{pyvx} and \eqref{quadrupole1}, which is shown in the upper panel of Fig.~\ref{fig:fig2}(c). This value aligns with the fact that the polarization of Wannier center, $p^{v_x}_y$ is quantized to 0 or $\pm0.5$~\cite{benalcazar2017electric} kept by the reflection $M_i$ symmetry. We turn to the second band gap and find it is still with quadrupole phase $q_{xy}=0.5$ because the fourth band adds nothing due to its trivial Chern and dipole phases, as shown in the lower panel of Fig.~\ref{fig:fig2}(c). As for the third gap, it has a non-trivial Chern phase with $C=1$ because the fifth and the sixth bands provide a non-trivial Chern phase $C=1$, illustrated by the bottom panel of Fig.~\ref{fig:fig2}(c).

Now in the case of $r=0.12a$, the first band gap is still a quadrupole gap while the second becomes a Chern gap. Also a new small gap in blue [cf. Fig.~\ref{fig:fig2} (a)] steps in with Chern phase $C=2$, which is absent in the case of $r=0.07a$. Furthermore, the fourth band gap is trivial in Chern, dipole, and quadrupole phases.

Such non-trivial topological invariants indicate that the edge and quadrupole states can cohabit in our Lieb lattice under open boundary conditions (OBC). To visualize this edge state, we construct a super cell consisting of $1\times 20$ unit cells, with vertical boundaries (in $y$ direction) set as PBCs [see inset in Fig.~\ref{fig:fig2}(d)]. To note, we set an air gap of width $g$ adjacent to the upper and lower boundaries of the super cell in panel (a) to push the edge and the corner states in the gap~\cite{he2020}. To be specific, the air gap depth $g$ serves to pull out the corner and edge states into the gap from the bulk~(see Append.~\ref{air gap} for further data). The projection band diagram of the super cell contains edge dispersion bands, as shown in red curves in Fig.~\ref{fig:fig2}(d). For the super cell consisting of 20 unit cells without any topological gaps, the number of bulk states under the first gap should be 60 because each unit cell under the first gap contributes 3 bands from Fig.~\ref{fig:fig2} (b). However, there are only 58 bulk states under the first band gap as shown in Fig.~\ref{fig:fig2} (d). The rest two states are then left in the first band gap. This count mismatch is a typical feature of higher-order topologies~\cite{vaidya2023,lan2024}. Moreover, a count mismatch by four edge states out of the bulk states below the second gap, due to its non-trivial quadrupole phase.




Similarly, for a super cell structure consisting of $10\times 10$ unit cells in Fig.~\ref{fig:fig2}(e) with four boundaries as OBCs, there are air gaps with $g=0.12a$ at the boundaries. Non-trivial quadrupole moments [presented in Fig.~\ref{fig:fig2} (c)] lead to topological corner states~\cite{he2020,benalcazar2017quantized}, as shown in Fig.~\ref{fig:fig2}(e). As mentioned above, the positions of the edge and the corner states caused by the quadrupole moment will change along with the varied air gap depth $g$. For example, we choose $g=0.12a$ intensionally to make the corner states more detached from the edge states, rather than $g=0.07a$ when the corner states in the second band gap are emerged in the bulk states [data shown in Fig.~\ref{fig:figA3}(a) in Append.~\ref{air gap}].


In this paragraph we now contrast the Chern, dipole, and quadrupole phases. In general, the Chern phase is considered of strong topology, which is intrinsic to the lattice and does not depend on the concrete structure of the unit cell, and its edge states are gapless, connecting the two parts of the bulk states. Differently, the dipole phase and quadrupole phase correspond to weak topology~\cite{fu2007, fu2007topological}, the edge states are usually gapped~\cite{lan2024}. And the position of the edge state caused by the dipole and quadrupole moment is not fixed in a specific band gap, but changes with the applied air gap depth and may even be pushed off into other band gaps~(cf. Append.~\ref{air gap}).

To sum up this section, the primitive Lieb PhC with uniform sublattices undergoes a well-defined phase transition as the sublattice radius increases. Specifically, it transitions from a pure quadrupole phase to a multiple phase regime along with two Chern phases ($C=1, 2$), where edge and corner states are characterized by the nontrivial quadrupole moment or Chern numbers.

\section{Topological phases in a deformed Lieb lattice: non-uniform radii for three sublattices}
\label{inhomogeneous}
\begin{figure*}[hbtp!]
\includegraphics[width=0.96\textwidth]{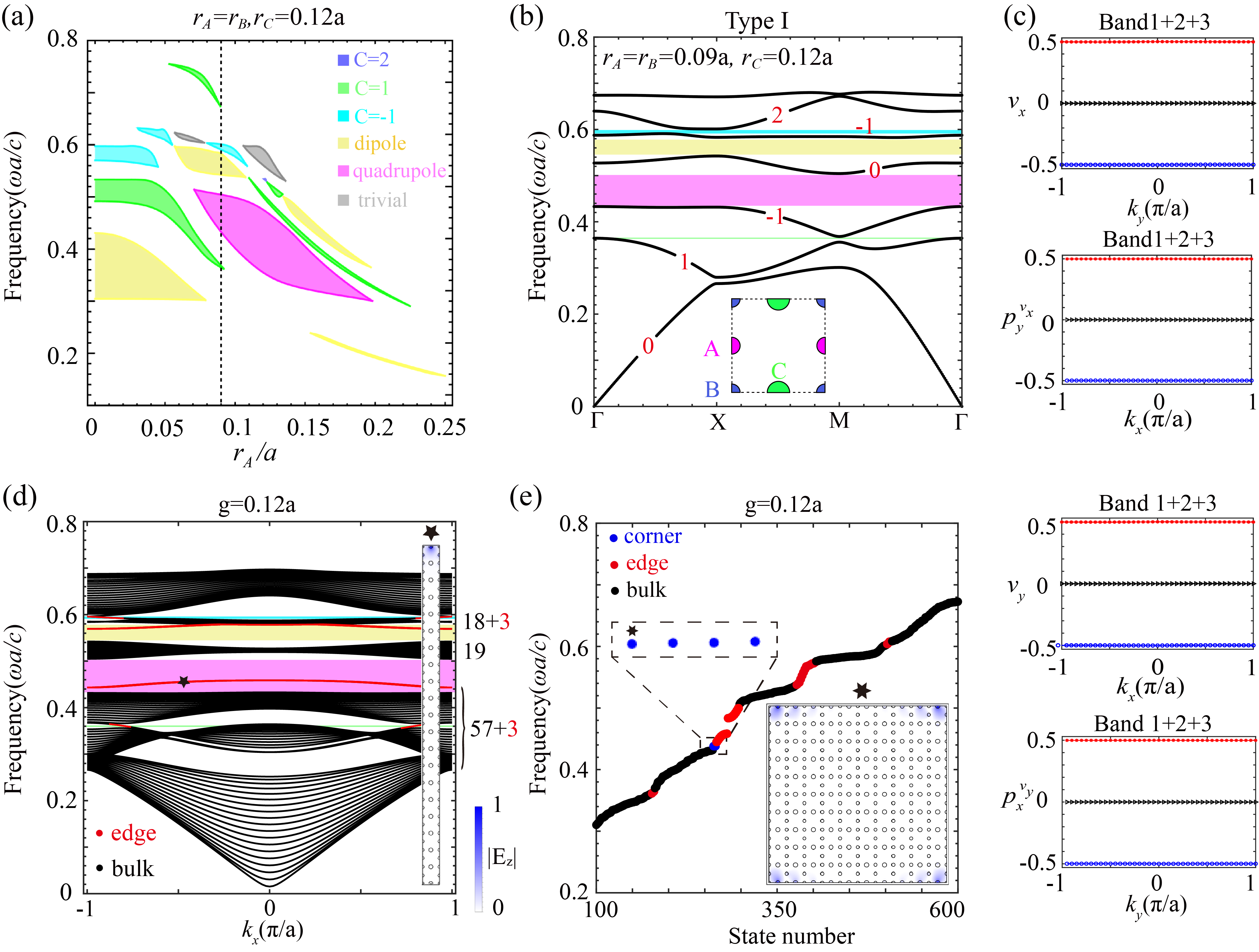}
\caption{\label{fig:fig3}(a) The band gap diagram with different topological phases as the radius $r_A$ increases. We fix $r_C=0.12a$ and request $r_A=r_B$ to introduce non-uniform radii. The coloration for each topological phase is indicated in its legend. (b) Band structure of such deformed Lieb PhC (type \uppercase\expandafter{\romannumeral1} lattice) with $r_A=r_B=0.09a$ and $r_C=0.12a$, with the Chern number of each band marked. The inset represents the unit cell structure, with each sublattice colored following panel (a). (c) The Wannier center $v_x$ and $v_y$ of the first three bands and their polarization $p_y^{v_x}$ and $p_x^{v_y}$. These Wannier centers indicate trivial dipole phases, and the polarization of the Wannier centers shows a non-trivial quadrupole phase. 
(d) Projected band diagram of the super cell consisting of $1 \times 20$ type \uppercase\expandafter{\romannumeral1} lattice. There are two air gaps with $g=0.12a$ at the top and bottom boundaries. The inset shows the electric field profile corresponding to the edge state marked by the pentagram. (e) Eigenstates of the super cell consisting of $10\times10$ type \uppercase\expandafter{\romannumeral1} lattice with the air gaps with $g=0.12a$ at the four boundaries. And the inset shows the electric field profile corresponding to the corner state marked by the hexagram. 
}
\end{figure*}

\begin{figure*}[htbp!]
\includegraphics[width=0.96\textwidth]{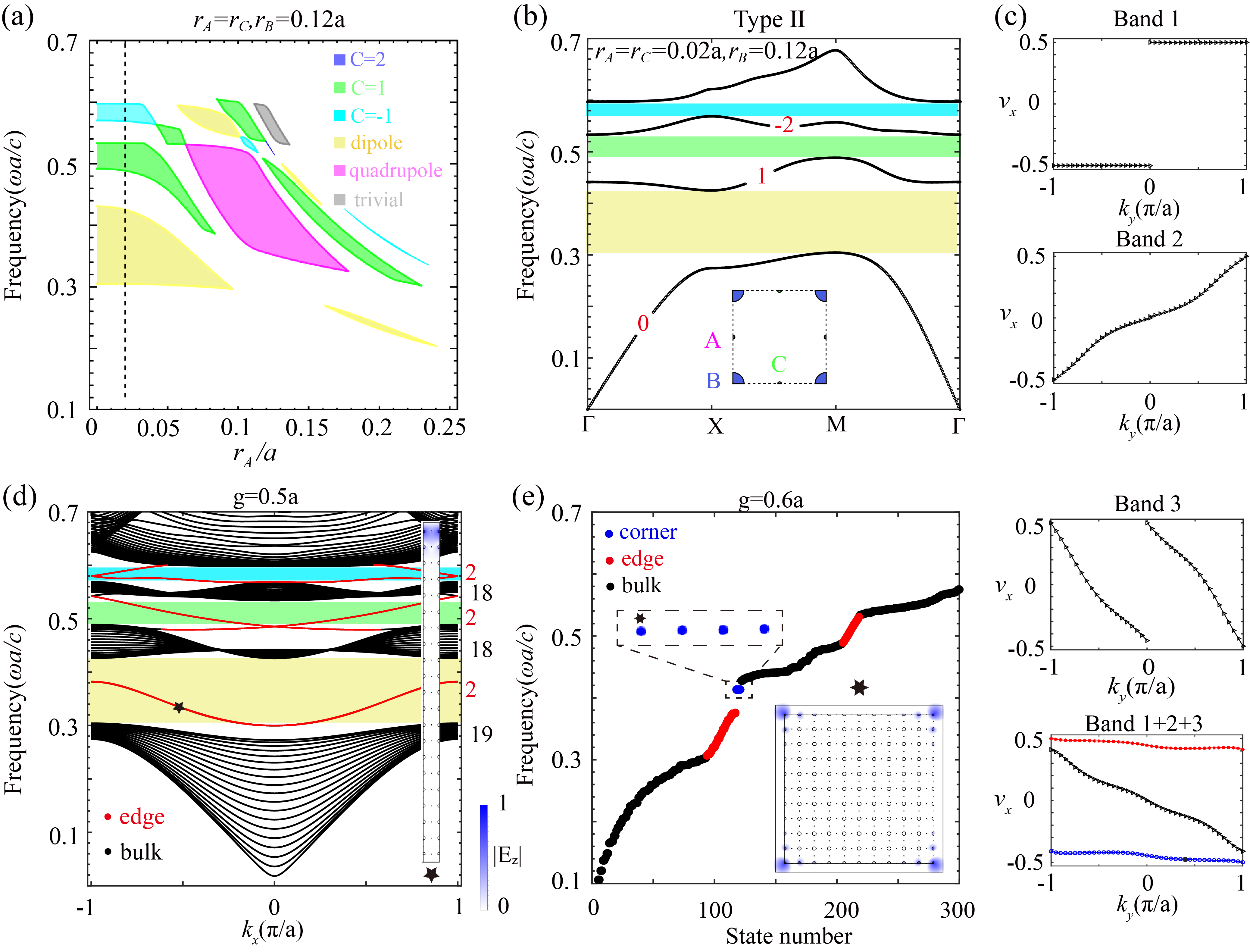}
\caption{\label{fig:fig4} (a) The band gap diagram with different topological phases as the radius $r_A$ increases. Here we fix $r_B=0.12a$ and define $r_A=r_C$ to introduce non-uniform radii. The coloration for each topological phase is indicated in its legend. (b) Band structure of type \uppercase\expandafter{\romannumeral2} lattice with radius $r_A=r_C=0.02a$ and $r_C=0.12a$, with the Chern number of each band marked. The inset represents the unit cell structure, and the coloured gaps indicate corresponding phases following panel (a). (c) The Wannier center $v_x$ distributions of each band. Top panel: the Wannier band of the first band, which shows non-trivial dipole phase; middle two panels: the Wannier band of the second and the third band, which show non-trivial Chern phases with $C=1$ and $C=-2$ respectively; Bottom panel: the Wannier band of the first three bands, which shows non-trivial Chern phase $C=-1$. 
(d) Projected band diagram of the super cell consisting of $1\times20$ type \uppercase\expandafter{\romannumeral2} lattices. There are two air gaps with $g=0.5a$ at the top and bottom boundaries respectively. The inset shows the electric field profile corresponding to the edge state marked by the pentagram. (e) Eigenstates of the super cell consisting of $10\times10$ type \uppercase\expandafter{\romannumeral2} lattices. The air gaps are $g=0.6a$ at the four boundaries and the inset shows the electric field profile corresponding to the corner state marked by the hexagram. 
}
\end{figure*}

In the previous Sec.~\ref{uniform}, we discuss the topological phase with uniform radii for Lieb lattice. In Sec.~\ref{inhomogeneous}, we then consider the deformation of the Lieb lattice with different radii $r_A, r_B$ and $r_C$ for the three sublattices, as shown in Fig.~\ref{fig:fig1}(c). Here we give two cases, one of which respectively breaks and maintains the $C_4$ symmetry of the unit cell, named as types \uppercase\expandafter{\romannumeral1} and \uppercase\expandafter{\romannumeral2} shown in Figs.~\ref{fig:fig3} and ~\ref{fig:fig4}.

Firstly, we consider the former case with broken $C_4$ symmetry: i.e., $r_A=r_B\neq r_C$, which still maintains the reflection symmetry $M_i$. Therefore, the dipole polarization is still quantized to $P_i=0$ or $0.5$, but may characterize unequal dipole moments $P_x$ and $P_y$. The diagram of band gaps in Fig.~\ref{fig:fig3}(a) demonstrates that as $r_A$ increases, the gaps shrink and generate one after another thus forming a series of topological phase transitions, among which a quadrupole gap (in magenta) dominates. Starting from $r_A=r_B=0$ as a square lattice, it shows non-trivial dipole ($P_x=0$, $P_y=0.5$) and Chern phases ($C=\pm1$). When $r_A$ gradually increases, the first dipole gap width with $P_y=0.5$ decreases until closed at $r_A=0.08a$, and the second gap with $C=+1$ closes at $r_A=0.09a$. Further increasing $r_A$ up to $0.06a$ causes another dipole gap higher than $0.52c/a$. When $0.07a<r_A<0.2a$, a quadrupole gap with $q_{xy}=1/2$ comes into play, occupying a large magenta portion in Fig.~\ref{fig:fig3}(a). For $r_A\geq 0.15a$, two new tiny dipole gaps appear, which sandwich the quadrupole gap just mentioned. The other four gaps with Chern phase also come around at various frequency ranges, shown as three green ribbons ($C=1$) and one blue spot ($C=2$) in Fig.~\ref{fig:fig3}(a).

Here, we select the case of $r_A=r_B=0.09a$ and name it as type \uppercase\expandafter{\romannumeral1} lattice, which is marked by the dashed line in Fig.~\ref{fig:fig3}(a) to pin down its topological features. For type \uppercase\expandafter{\romannumeral1} lattice, its band structure shown in Fig.~\ref{fig:fig3}(b) reveals four topological gaps with Chern, dipole and quadrupole phases marked following the legend in Fig.~\ref{fig:fig3}(a) along with the Chern numbers for each band. For the first tiny gap in green [cf. panel (b)], it carries a Chern number $C=1$ due to the second band below it. And the second gap in magenta carries non-trivial quadrupole phase. The Wannier centers of the first three bands $v_{i}(i=x,y)$ show trivial dipole phase while their polarizations show non-trivial quadrupole moment $q_{xy}=0.5$ from Eqs.~\eqref{pyvx} and \eqref{quadrupole1}, as all shown in Fig.~\ref{fig:fig3}(c). The third gap carries non-trivial dipole phase with $P_y=0.5$ due to the fourth band. And finally the fourth gap is of Chern phase $C=-1$.

To reveal the edge and corner states, we construct a super cell of type \uppercase\expandafter{\romannumeral1} lattice consisting of $1 \times 20$ unit cells with two air gaps $g=0.12a$ at both boundaries. The projected diagram of such a super cell shown in Fig.~\ref{fig:fig3}(d) contains the tiny first gap barely legible and three other gaps containing several edge states. The inset shows the electric field profile for the edge state marked by the pentagram. When we set all four boundaries as OBCs, corner states appear in the second gap with quadrupole phase, as shown in Fig.~\ref{fig:fig3}(e) and its inset for the corner state marked by the hexagram.

Secondly, we consider the latter case with $C_4$ symmetry: $r_A=r_C, r_B=0.12a$. Figure~\ref{fig:fig4}(a) reveals rich topological phases of gaps as $r_A$ increases, similar to Fig.~\ref{fig:fig3}(a). When $r_A=r_C=0$, the structure is a square lattice~\cite{lan2024} of non-trivial dipole $P_x=P_y=0.5$ and Chern phases $C=\pm1$. When $r_A$ gradually increases to $0.062a$, the first three gaps decrease and finally close. Also the third gap flips from $C=1$ to $C=-1$ at about $0.046a$. Again a major area of quadrupole phase with $q_{xy}=0.5$ occurs within the interval of $0.062a<r_A<0.178a$, which seems to be the feature for primitive and deformed Lieb lattices. Other tiny gaps of Chern and dipole phases also appear at high frequency regions.

For example, we choose $r_A=r_C=0.02a$ and name it as type \uppercase\expandafter{\romannumeral2} lattice to give more details about its topological phases, which is marked by the dashed line in Fig.~\ref{fig:fig4}(a). The band structure of the type \uppercase\expandafter{\romannumeral2} lattice, shown in Fig.~\ref{fig:fig4}(b), occupy three band gaps: the first gap shows non-trivial dipole phase for $P_x=P_y=0.5$, as shown in the top panel of Fig.~\ref{fig:fig4}(c); the second gap is of Chern phase $C=1$, as shown by its Wannier center in the upper panel of Fig.~\ref{fig:fig4}(c) (a positive winding number $C_2=1$). And the Wannier center of the third band gives $C_3=-2$, and the Chern number for the third gap is then $C=-1$,as shown in the lower and the bottom panels of Fig.~\ref{fig:fig4}(c), respectively.

The non-trivial topological phases can lead to edge or even corner states under OBCs. We construct a super cell consisting of $1\times20$ unit cells, and set an air gap with width $g=0.5a$ adjacent to its upper and lower boundaries. The projected band of the super cell shown in Fig.~\ref{fig:fig4}(d) contains several edge states, which position in the first gap with dipole phase, in the second with Chern phase ($C=1$), and in the third with Chern phase ($C=-1$). The inset profile shows the edge state marked by the pentagram in the band diagram. And for a super cell structure consisting of $10\times10$ cells with an air gap depth of $g=0.6a$, four corner states appear in the first gap, as shown in the frequency diagram of Fig.~\ref{fig:fig4}(e), and its electric field as inset assures its corner nature. Also, the position of edge and corner states caused by non-trivial dipole polarization can be adjusted via changing the air gap depth [see Fig.~\ref{fig:figA3}(b) in Append.~\ref{air gap}].

In this section, we induce mottled topological phases in the deformed Lieb lattice by varying two radii of the three sublattices, which are richer than those of the primitive lattice. Notably for such a deformed lattice, a nontrivial quadrupole band gap persists, whether the $C_4$ symmetry of unit cell is broken or not.

\section{Topological phases in a deformed Lieb lattice: shifting sublattices in the unit cell}\label{shift}
In Secs.~\ref{uniform} and~\ref{inhomogeneous} above, we obtain topological phases by tuning the radii of the sublattices uniformly and non-uniformly, respectively. In this section, we will explore another tuning degree of freedom for unit cells, i.e., shifting the sublattices, to acquire further topological phase transitions. As shown in Fig.~\ref{fig:fig1}(d), parameters $\delta d_{j} (j=A, B, C)$ represent the distances for the sublattice $j$ to deviate from its original position in the primitive lattice. In Sec.~\ref{shift}, we will use two examples to demonstrate the phase transition via shifting sublattices: the radii of sublattices  are set as uniform $r=r_A=r_B=r_C=0.12a$ and only sublattices A and C are shifted (sublattice B fixed, $\delta d_B=0$). We will find out that shifting the sublattices will destroy the reflection symmetry of the lattice, causing the system to acquire a non-quantized dipole or quadrupole moment. 

\begin{figure*}[hbtp!] 
\includegraphics[width=0.96\textwidth]{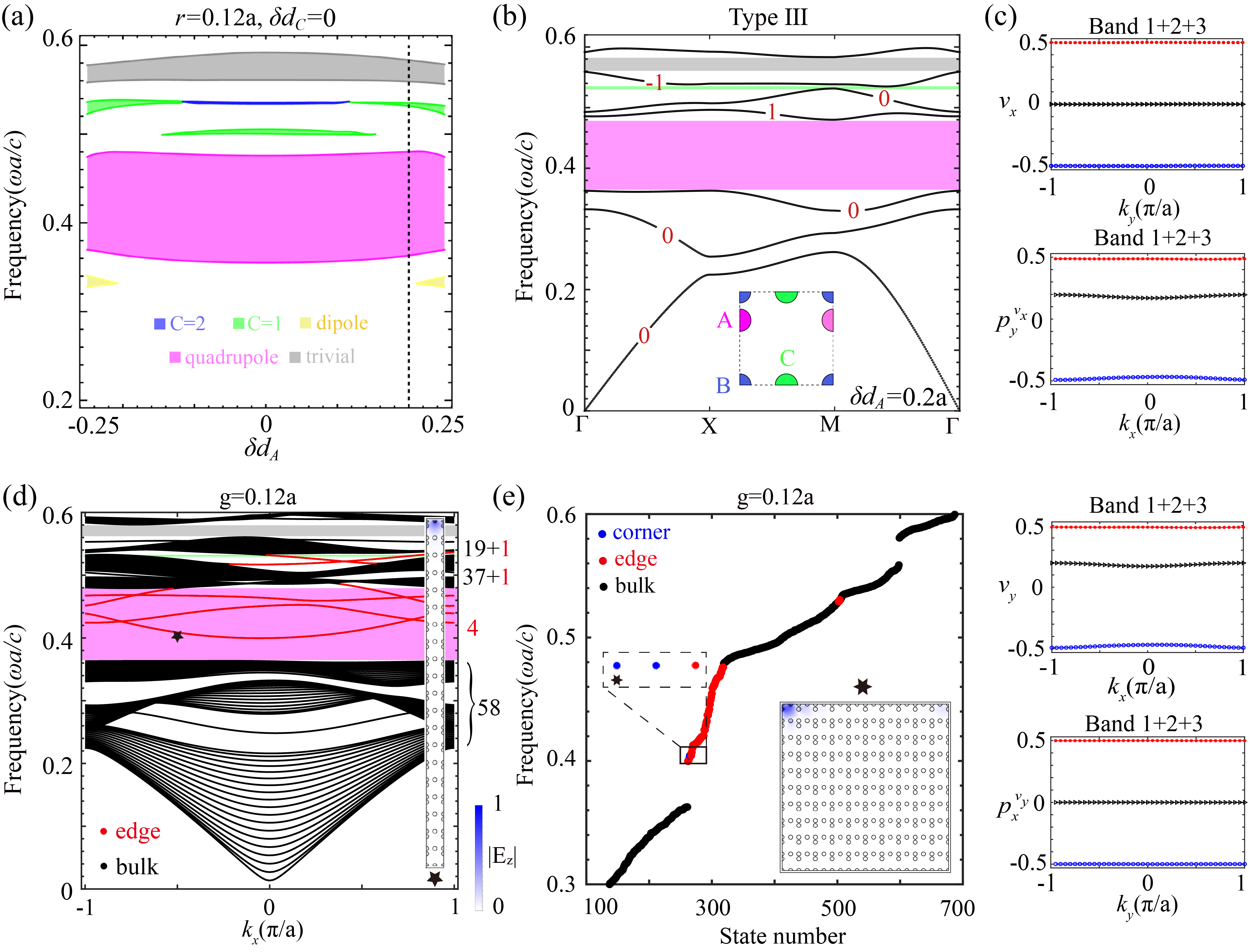}
\caption{\label{fig:fig5} (a) The band gap diagram with different topological phases as the shifted distance $\delta d_A$ increases. Here we fix the position of sublattice C, i.e. $\delta d_C=0$ and define $r_A=r_B=r_C=0.12a$. The coloration for each topological phase is indicated in its legend. (b) Band structure of the type \uppercase\expandafter{\romannumeral3} lattice with $\delta d_A=0.2a$, with the Chern number of each band marked. The inset represents the unit cell and the coloured gaps indicate corresponding phases following panel (a). (c) The Wannier center $v_x$ and $v_y$ distributions of the first three bands and their polarization $p_y^{v_x}$ and $p_x^{v_y}$. 
(d) Projected band diagram of the super cell consisting of $1\times20$ type \uppercase\expandafter{\romannumeral3} lattices, with $g=0.12a$ at the top and bottom boundaries. The inset shows the electric field profile corresponding to the edge state marked by the pentagram. (e) Eigenstates of the super cell consisting of $10\times10$ type \uppercase\expandafter{\romannumeral3} lattices, with air gaps $g=0.12a$ at the four boundaries and the inset for the electric field profile of the corner state marked by the hexagram. 
}
\end{figure*}

Firstly, we only shift the position of sublattice A (or C, since the two cases are equivalent), which will break $M_y$ but maintain $M_x$ reflection symmetry. The band gap diagram as a function of $\delta d_A$ in Fig.~\ref{fig:fig5}(a) appears symmetric and also rich. When $\delta d_A=0$, the structure is just the primitive lattice with $r=0.12a$, whose phases of quadrupole $q_{xy}=0.5$ and Chern phases $C=1, 2$ are shown in Fig.~\ref{fig:fig2}(a). As one shifts sublattice A, the first band gap width remains almost unchanged. In other words, there is no closing and opening of the band gap during the movement of sublattice A. And the second gap in green with $C=1$ will close when $\vert \delta d_A\vert >0.15a$. The third gap with Chern phase $C=2$ remains when $|\delta d_A|<0.11a$. When one increases $\vert \delta d_A\vert$ further, the gap flips to another Chern phase $C=1$. Moreover, when $|\delta d_A|>0.21a$, there is an additional gap beneath with non-zero dipole polarization $P_y=0.41$ and $P_x=0.5$.

Specifically, we select a concrete case $\delta d_A=0.2a, \delta d_C=0$, named as type  \uppercase\expandafter{\romannumeral3} lattice, which is marked by the dashed line in Fig.~\ref{fig:fig5}(a). The band structure of type \uppercase\expandafter{\romannumeral3} lattice in Fig.~\ref{fig:fig5}(b) demonstrates three gaps, of which the first band gap holds a quadrupole phase. Firstly, one obtains $P_x=0$ and $P_y=0.19$ via the Wannier center of the first three bands $v_x$ and $v_y$. Due to the broken $M_y$ reflection symmetry, the distribution of Wannier center $v_y$ is not quantized to 0 or 0.5. Then the polarizations of Wannier center $v_x$ and $v_y$ are calculated to arrive at $q_{xy}=0.47$~[cf. Fig.~\ref{fig:fig5}(c)]. This quadrupole polarization results from a quantized $p_x^{v_y}$ under $M_x$ symmetry and a non-quantized $p_y^{v_x}$ for broken $M_y$ symmetry~\cite{benalcazar2017electric}. Now we construct the super cell consisting of $1\times 20$ units for  type \uppercase\expandafter{\romannumeral3} lattice with an adjacent air gap $g=0.12a$. The projected band diagram shown in Fig.~\ref{fig:fig5}(d) contains four edge states in the first gap with count mismatch. Moreover, when we set both $x$ and $y$ boundaries to OBCs, corner states appear in the first gap with a quadrupole phase, as shown in Fig.~\ref{fig:fig5}(e).

\begin{figure*}[hbtp!] 
\includegraphics[width=0.94\textwidth]{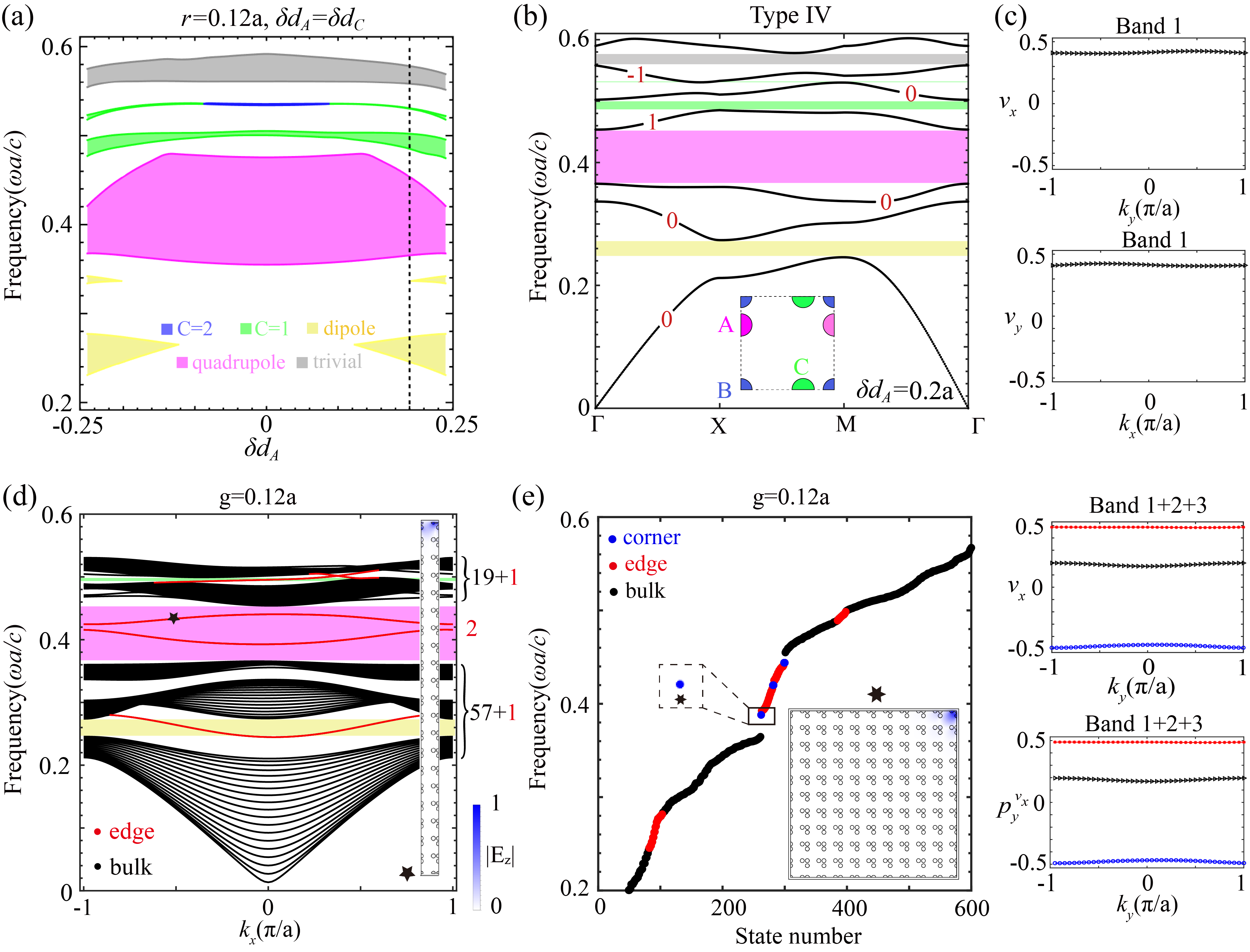}
\caption{\label{fig:fig6} (a) The diagram of the band gaps as the shifted distances of sublattices A and C increase, i.e., $\delta d_A=\delta d_C$. Here we still define $r_A=r_B=r_C=0.12a$. The coloration for each topological phase is indicated in its legend. (b) Band structure of the type \uppercase\expandafter{\romannumeral4} lattice with $\delta d_A=\delta d_C=0.2a$, with the Chern number of each band marked. The inset represents the unit cell structure, and the coloured gaps indicate corresponding phases following panel (a). (c) The Wannier center $v_x$ and $v_y$ distributions of the first three bands and their polarization $p_y^{v_x}$ and $p_x^{v_y}$. 
(d) Projected band diagram of the super cell consisting of $1\times20$ type \uppercase\expandafter{\romannumeral4} lattices, with the air gap $g=0.12a$ at the top and bottom boundaries. The inset shows the electric field profile corresponding to the edge state marked by the pentagram. (e) Eigenstates of the super cell consisting of $10\times10$ type \uppercase\expandafter{\romannumeral4} lattices. The air gaps with $g=0.12a$ at the four boundaries and the inset shows the electric field profile corresponding to the corner state marked by the hexagram. }
\end{figure*}

Secondly, we move sublattices A and C synchronously, i.e., $\delta d_A=\delta d_C$, which will break the $M_x$ and $M_y$ reflection symmetries for such a deformed lattice, resulting in non-quantized dipole and quadrupole phases. The frequency diagram is charted in Fig.~\ref{fig:fig6}(a): only the third band gap undergoes a topological phase transition from $C=2$ to $C=1$ at $\delta d_A=\delta d_C=\pm0.09a$, and all the other gaps retain their topological phases within the calculated range of $\vert\delta d_A\vert<0.25a$. Specifically, we select the case of $\delta d_A=\delta d_C=0.2a$ and name the structure as type \uppercase\expandafter{\romannumeral4} lattice, which is marked by the dashed line in Fig.~\ref{fig:fig6}(a). The band structure of type \uppercase\expandafter{\romannumeral4} lattice shown in Fig.~\ref{fig:fig6}(b) shows three band gaps. For the first gap, the Wannier centers $v_x$ and $v_y$ result in dipole moment $P_x=P_y=0.41$, shown in the top and upper panels of Fig.~\ref{fig:fig6}(c). For the second gap, the Wannier center $v_x$ of the first three bands, as shown in the lower panel of Fig.~\ref{fig:fig6}(c). Interestingly, the polarization of $v_x$, as shown in the bottom panel of Fig.~\ref{fig:fig6}(c), is the same as the polarization of $v_y$, which leads to $q_{xy}=0.48$. The projected band shown in Fig.~\ref{fig:fig6}(d) contains edge states present in all the three gaps. When both $x$ and $y$ boundaries are set to OBCs, a corner state in the second gap occurs for the quadrupole phase, as shown in Fig.~\ref{fig:fig6}(e).

To close this section, shifting the distances between sublattices to break the reflection symmetry of the unit cell results in non-quantized polarizations of dipole and quadrupole phases.

\section{Further remarks} \label{remark}

To give some further insights for possible experimental setups from our built numeric results of the displaced Lieb PhC lattices, three remarks are listed here on non-quantized moments, a phase diagram for deformation parameters and the magneto-optical coupling strength.

\subsection{Less robustness for non-quantized moments and experimental challenges}
\begin{figure*}[htp!] 
\includegraphics[width=0.84\textwidth]{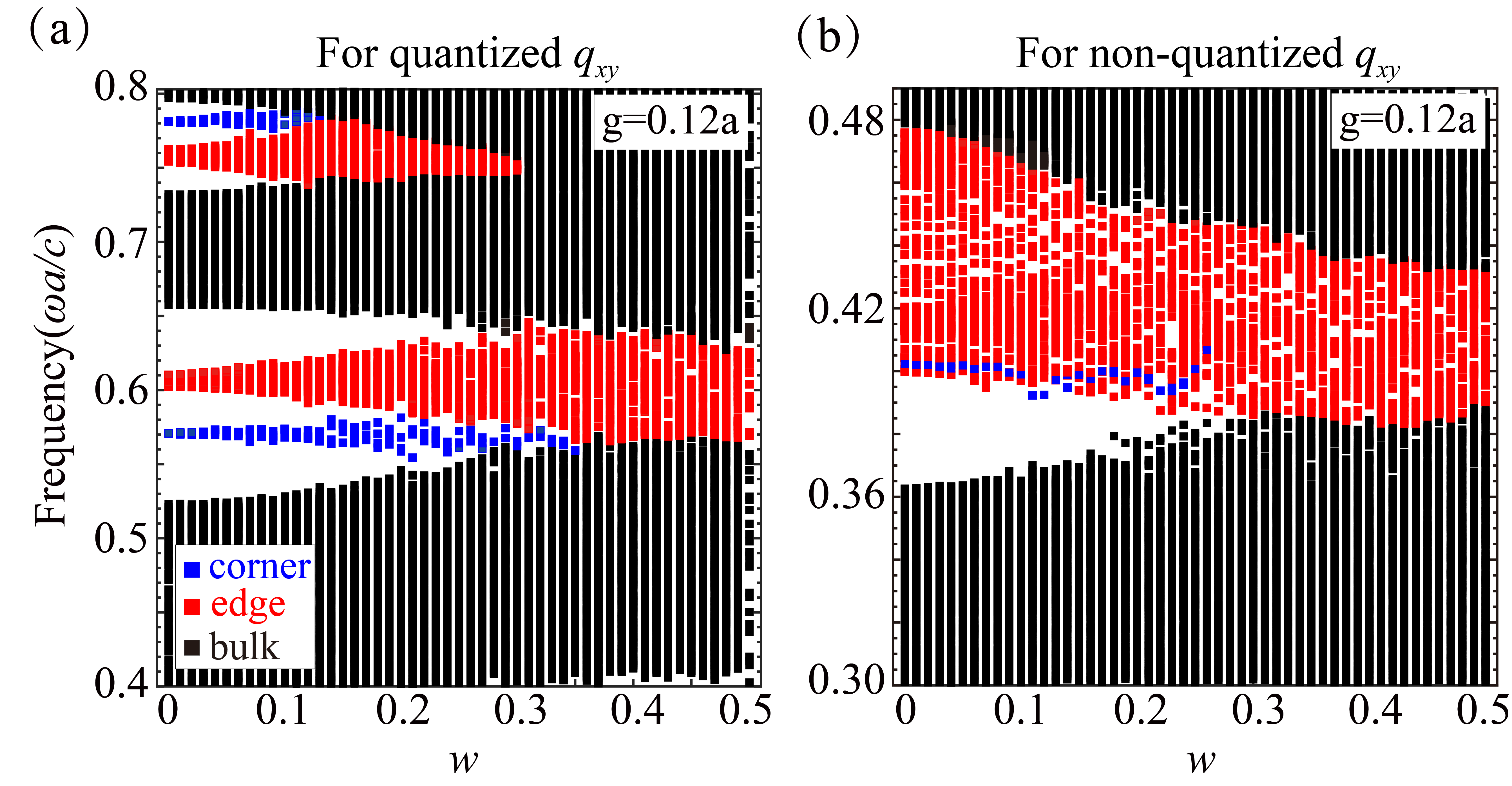}
\caption{\label{fig:fig7}State frequencies vary under disorder strength $w$: (a) corner and edge states in the first and second gaps for Fig.~\ref{fig:fig2}; (b) those in the first gap for Fig.~\ref{fig:fig5}. The square markers represent corner states (blue), edge states (red), and bulk states (black). Each sublattice radius $r$ is perturbed by a uniform disorder $\Delta r=w\xi$ ($\xi\in[-0.5, 0.5]$). } 

\end{figure*}

For sublattice shifting in Fig.~\ref{fig:fig5}, non-quantized dipole/quadrupole moments are observed in panel (c), which can be attributed to reflection symmetry breaking ($M_x$, $M_y$ or $C_4$) on the Wannier centers or Wilson loops (see~Append.~\ref{symmetry}). Nevertheless, it also signifies a bulk polarization for dipole or quadrupole moments, albeit not symmetry-enforced. Thus it also differs in terms of state stability and disorder resistance from quantized values. To elucidate the physical significance of the non-quantized moments in Fig.~\ref{fig:fig5}(c), we contrast in Fig.~\ref{fig:fig7} the state frequency variances under disorder below $w=0.5$ for two scenarios: (a) quantized moments for primitive lattices in Fig~\ref{fig:fig2}, and (b) non-quantized ones for type \uppercase\expandafter{\romannumeral3} in Fig.~\ref{fig:fig5}. For both lattices, all radii $r$ for sublattices are perturbed by random disorder
\begin{equation}
\Delta r=w\xi, \quad \xi\in[-0.5, 0.5],
\end{equation}
where for each radius $\xi$ is a random variable obeying uniform distribution. Both panels in Figs.~\ref{fig:fig7} reveal a clear gap closing when the disorder strength $w$ reaches certain thresholds. To be specific, (a) for quantized moments the higher-order states diminish for threshold $w=0.35$ while (b) for non-quantized ones they do so for $w=0.25$. Such a robustness difference of the corner and edge states against disorder then leads us to speculate that non-quantized moments indicate less robustness of higher-order topological states than the quantized counterparts. And the corner states are immersed in the edge ones for non-quantized moments [also occurring in~Figs.~2(e)-6(e)], which may be a feature worth looking at in future works.  

We now note about technical challenges for our theoretical proposals in experimental implementation. Possible technical bottlenecks with existing microfabrication could be summarized as follows. (1) One is the difficulty to maintain our PhC structure to obey the exact spatial symmetry as designed. Current electron-beam lithography and dry etching processes, when applied to YIG, will suffer from disorder of manufacture errors~[cf.~Fig.~\ref{fig:fig7}], which could obscure the theoretically-predicted gap and make it difficult to distinguish corner states from disorder-induced localized modes. (2) The second can be the difficulty to realise a uniform bias magnetic field to break time-reversal symmetry. This issue may deviate experimentalist from calibrating high-resolution real-space imaging. (3) Finally, a PEC boundary with an air gap in our model corresponds to a metal cladding in experiments. This metal cladding however, may result in significant loss at optical frequencies to perturb the fragile topological boundary state off from our predictions.

\subsection{A phase diagram for bandgap widths on deformation parameters of unit cell}
\begin{figure*}[htp!] 
\includegraphics[width=0.7\textwidth]{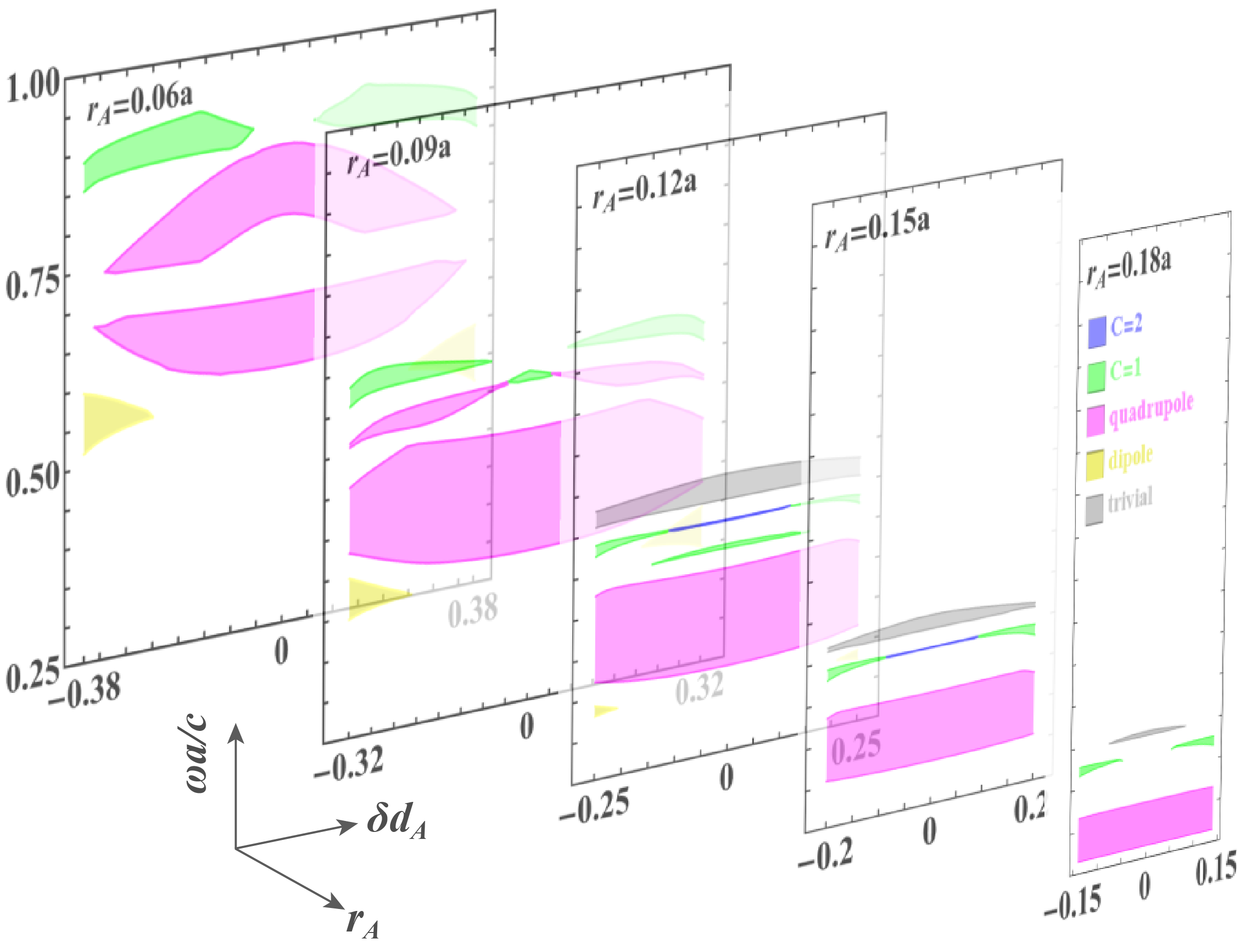}
\caption{\label{fig:fig8}Phase diagram in the deformation space $(r_A, \delta d_A)$ including frequency diagrams for five radii of sublattices equally displaced for design benchmark: $r_A/a=0.06,0.09,0.12,0.15,$ and $0.18$. Here, we assume all the pillars with uniform radii, i.e., $r=r_A=r_B=r_C$ (using parameters for Fig.~\ref{fig:fig5}). The five panels are arranged horizontally with increasing $r_A$ from left to right. In each panel, the horizontal axis represents $\delta d_A$, quantifying the symmetry-breaking shift of sublattice A. Distinct topological phases are delineated by colored regions following its legend of the panel of $r_A=0.18a$.} 
\end{figure*}

To quantify further how the bandgap widths depend on deformation parameters, a new phase diagram of gaps depend on $(r_A, \delta d_A)$ is presented in Fig.~\ref{fig:fig8} for five discrete radii of sublattices (using parameters for Fig.~\ref{fig:fig5}). As the radius for sublattice A  increases from $r_A=0.06a$ to $0.18a$, three facts are observed in the numeric data. (1) All bandgaps tend to lower and shrink monotonically as the sublattice radius increases. This means that the more sublattice A shifts to break the spatial symmetry of unit cells, the more likely the gaps tend to close up and lower down simultaneously. (2) The range of $\delta r_A$ also tends to shrink as the radius shifts further because the sublattices touch each other just at the limited ends for $\delta r_A$. (3) The phase diagram is symmetrical along the axis of $\delta r_A=0$. 

For the least radius $r_A=0.06a$, four band gaps occur: a Chern gap with $C=1$, two quadrupole, and one dipole one. As $r_A$ increases to $0.09a$ the topological gaps shrink downwards until at $0.12a$ where the gaps of $C=2$ occur just sandwiched by the $C=1$ gaps under the trivial gap. Then for the largest radius $r_A=0.18a$, the $C=2$ gap already vanishes and the rest three gaps remain. These concurrent multiple phases with a limited data range infer that they are within possible fine tuning ranges in our Lieb PhC. The data visualization of Fig.~\ref{fig:fig8} maps out a glimpse of bandgap widths with respect to the geometric parameters $(r_A, \delta d_A)$ and could serve as a design guide for accessing targeted topological states in experiment.

\subsection{Magneto-optical coupling strength affects the quadrupole gaps}
	It remains to elaborate how the magneto-optical coupling strength $\kappa$ in Eq.~\eqref{0} affects the stability and phase boundaries of the higher-order gaps, particularly the quadrupole ones. We shall remark about this point as follows. 

In our numeric cases, the gyromagnetic material (YIG) possesses an off-diagonal term $\kappa$, which breaks time-reversal symmetry (TRS) and allows non-zero Chern numbers to emerge, i.e., first-order topological phases. For higher-order topological phases such as the quadrupole phase, their stability also relies on the spatial symmetries such as reflection symmetries $M_x$ and $M_y$ and rotation symmetry $C_4$. The magneto-optical effect itself does not directly break these spatial symmetries; rather, by breaking TRS it enables the band structure to generate new higher-order gaps that are absent in dielectric systems [e.g., the third band gap in Fig.~\ref{fig:fig2}(a)]. Thus the role of the magneto-optical effect is to create richer band gaps for higher-order topological phases across a broader frequency spectrum. 

\begin{figure}[htp!] 
\includegraphics[width=0.42\textwidth]{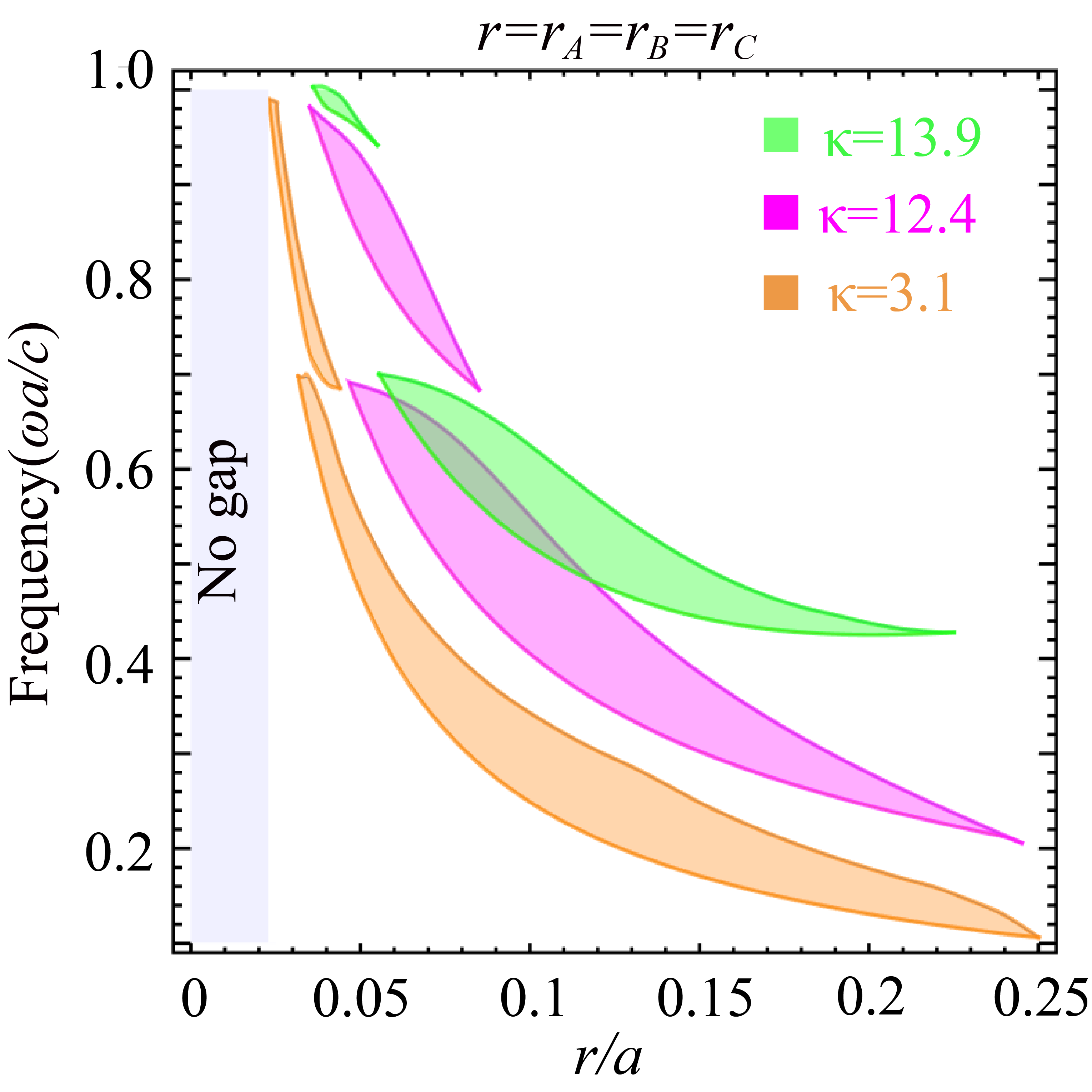}
\caption{\label{fig:fig9}
Frequency diagram for quadrupole gaps for the primitive lattice in Fig.~\ref{fig:fig2} when the magneto-optical coupling strength $\kappa$ varies. The gaps above $\kappa=13.9$ are not presented because new zero-energy states and negative determinant of permeability tensor occur therein: $\det \bar{\bar{\mu}}<0$. }  

\end{figure}

To be precise, the topological phase boundaries (such as the opening and closing of quadrupole band gaps) are determined by the modulation of the unit cell symmetry through geometric parameters (such as $r$ and $\delta d$), and by the coupling strength $\kappa$ both defined in Eq.~\eqref{0}. The magneto-optical effect twists the overall topology of the bands by breaking time-reversal symmetry, thereby affecting higher-order topological phases. We vary the strength of the magneto-optical effect-$\kappa$ to showcase its influence on topological band gaps and phase boundaries in Fig.~\ref{fig:fig9}, which shows a trend to shrink the quadrupole gaps as the coupling strength increases. This line shall represent another promising direction for our future works.

\section{Conclusion}

In summary, we map out the topological phases of a gyromagnetic PhC in the Lieb lattice with primitive and deformed structures. These lattices host Chern, dipole, and quadrupole topological states enriched by the broken time-reversal symmetry, which is confirmed by calculating higher-order polarizations. For the primitive gyromagnetic Lieb lattice, it shows both phases of Chern and quadrupole. And for a deformed Lieb lattice with unequal sublattice radii, a dipole band gap arises; for one with shifted sublattices, even richer topological phases are induced, with dipole moments and higher Chern numbers~\cite{tian2023breakdown}.

Our examples establish the sublattice radii and the positional shifts as two tunable parameters driving topological phase transitions in the gyromagnetic PhC platform. And the learnt knowledge from Lieb lattices shall apply well to other types of periodic structures~\cite{schindler2018, kunst2019boundaries, ZhangY2025Topo_Corner}, whose underlying principles in our work--combining broken time-reversal symmetry (TRS) with controlled geometric deformation--can well apply to similar topological hierarchies in other three-sublattice structures, such as Kagome~\cite{Bolens2019breathing_Kagome} or dice lattices~\cite{WangF2011dice, Mohanta2023Majorana}. So our work will contribute to the arsenal for versatile topological phases which are readily extendable to other concrete instruments~\cite{zhang2020symmetry, wei2025higher}, such as acoustic devices and electrical circuits to exploit this abstract idea in topology.

\appendix %

\section{Chern number, dipole moment and quadrupole moment}\label{Method}
\renewcommand{\thefigure}{A\arabic{figure}} 
\setcounter{figure}{0} 

\begin{figure}[hbtp!]
\includegraphics[width=0.45\textwidth]{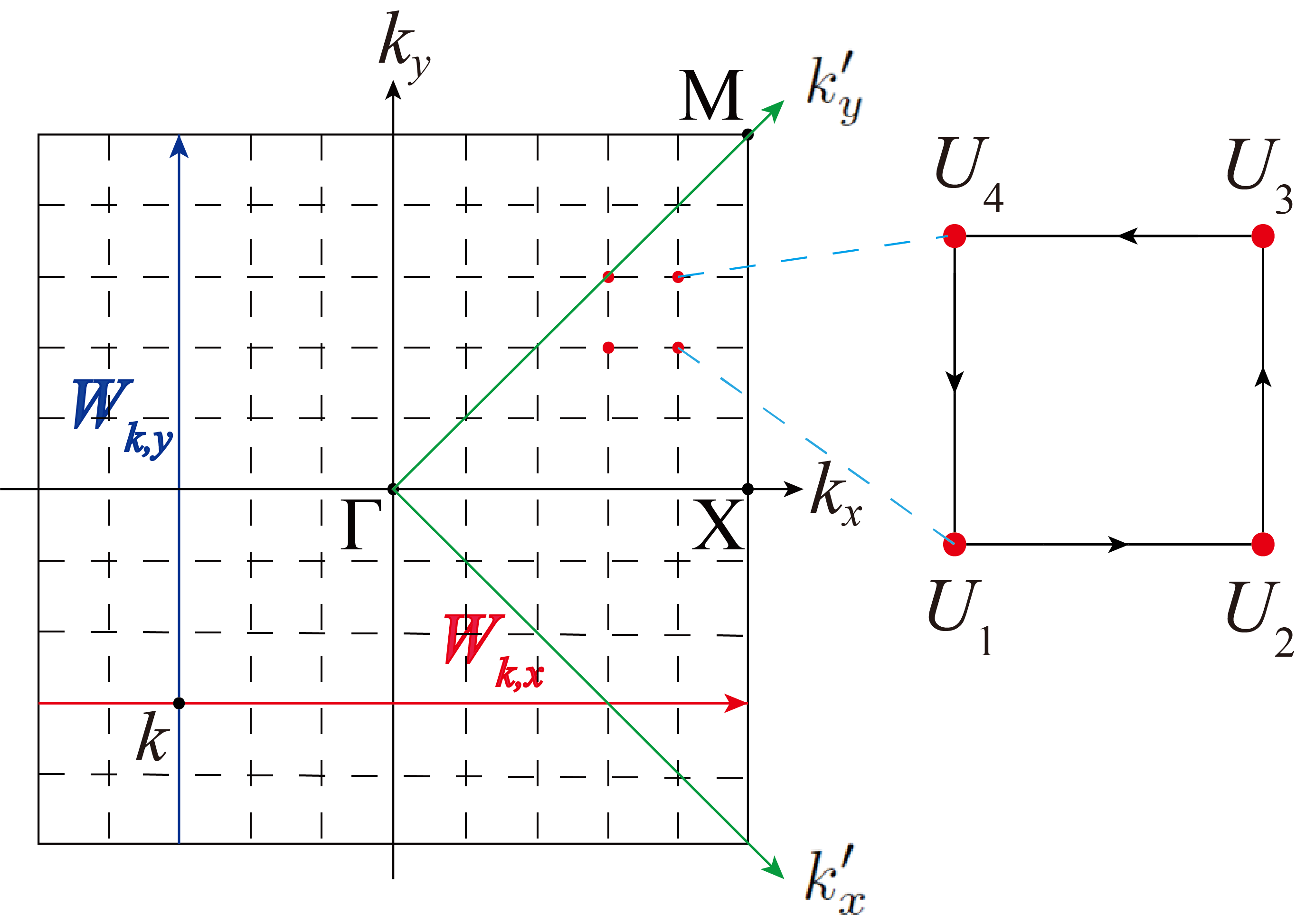}
\caption{\label{fig:figA1} The discretized first BZ, solving the Berry curvature of each small square and summing them up, we can get the Berry curvature of the entire first BZ. Here we introduce in detail the calculation method of Berry curvature of one of the small squares. From the base point $\textbf{k}(k_x, k_y)$, the red and blue arrows represent the Wilson loop in the first BZ along the $k_x$ and $k_y$ directions, respectively. In addition, we can rotate the axes by $\pi/4$ to obtain new coordinates, and the axes of new coordinates are marked as $k'_x$ and $k'_y$. 
}
\end{figure}

Firstly, we use the discretized BZ method to calculate the Berry curvature of each square unit, as shown in Fig.~\ref{fig:figA1}. The Chern number for band $n$ is presented as
\begin{eqnarray}
C_n=\frac{1}{2\pi i}\iint_{BZ}dS_{\mathbf{k}}\hat{z}\cdot\mathbf{F}^n(\textbf{k}), 
\end{eqnarray}
where
\begin{eqnarray}
\textbf{F}^n(\textbf{k})=\nabla_\mathbf{k}\times \mathbf{A}^n(\mathbf{k}), \label{vecF}
\end{eqnarray}
is the Berry curvature, and
\begin{eqnarray}
\textbf{A}^n(\textbf{k})=i\langle u^n_{\textbf{k}} \vert \nabla_{\textbf{k}} \vert u^n_{\textbf{k}}\rangle, \label{Ak}
\end{eqnarray}
is the Berry connection. And the TM eigenmode for band $n$ is defined as Bloch function 
\begin{equation}
\mathbf{E}_k^n(\mathbf{r})=\hat{z}E_k^n(\mathbf{r})=\hat{z}u^n_{\mathbf{k}}(\mathbf{r}) e^{i\mathbf{k}\cdot \mathbf{r}}. 
\end{equation}

Here we select a small part of the closed loop $U_{1\rightarrow 4}$. By calculating the curvature around the loop and integrating it in the first BZ, the total Berry curvature can be obtained, thereby obtaining the Chern number of band $n$~\cite{wang2020,goudarzi2022, tian2023breakdown}, as
\begin{eqnarray}
C_n=\frac{1}{2\pi i}\sum_{\textbf{k}_l\in BZ}\ln(U_{1,2}U_{2,3}U_{3,4}U_{4,1}),
\end{eqnarray}
with
\begin{eqnarray}
U_{1,2}(\textbf{k}_l)&=&\frac{\langle u(\textbf{k}_l)|u(\textbf{k}_l+\delta\textbf{k}_x)\rangle}{|\langle u(\textbf{k}_l)|u(\textbf{k}_l+\delta\textbf{k}_x)\rangle|},\nonumber\\
U_{2,3}(\textbf{k}_l)&=&\frac{\langle u(\textbf{k}_l+\delta\textbf{k}_x)|u(\textbf{k}_l+\delta\textbf{k}_x+\delta\textbf{k}_y)\rangle}{|\langle u(\textbf{k}_l+\delta\textbf{k}_x)|u(\textbf{k}_l+\delta\textbf{k}_x+\delta\textbf{k}_y)\rangle|},\nonumber\\
U_{3,4}(\textbf{k}_l)&=&\frac{\langle u(\textbf{k}_l+\delta\textbf{k}_x+\delta\textbf{k}_y)|u(\textbf{k}_l+\delta\textbf{k}_y)\rangle}{|\langle u(\textbf{k}_l+\delta\textbf{k}_x+\delta\textbf{k}_y)|u(\textbf{k}_l+\delta\textbf{k}_y)\rangle|},\nonumber\\
U_{4,1}(\textbf{k}_l)&=&\frac{\langle u(\textbf{k}_l+\delta\textbf{k}_y)|u(\textbf{k}_l)\rangle}{|\langle u(\textbf{k}_l+\delta\textbf{k}_y)|u(\textbf{k}_l)\rangle|}.
\label{U41}
\end{eqnarray}

Due to the $C_4$ rotation symmetry of the Lieb PhC, the dipole moment fulfill $P_x=P_y$, we focus on the $x$-component of the polarization $P_x$ in the following text~\cite{lan2023}, which is calculated from the eigenvalues of the Wilson loop $W_{k,x}$ along the $x$ direction, as shown in Fig.~\ref{fig:figA1}. One can define the Wilson line element from the base point $\textbf{k}=(k_x,k_y)$ to $\textbf{k}+\vartriangle_x$ along the $x$ direction as
\begin{eqnarray}\label{Fmn}
F^{mn}_{x,\textbf{k}}&=&\langle u^m_{\textbf{k}+\vartriangle_x}|u^n_{\textbf{k}}\rangle\nonumber\\
&=&\int_{\rm{unit cell}}dxdy\quad u^{m*}_{\textbf{k}+\vartriangle_x}(\textbf{r})\epsilon(\textbf{r})u^n_{\textbf{k}}(\textbf{r}),
\end{eqnarray}
where $|u^n_{\textbf{k}}\rangle$ is the Bloch wave function, which fulfills $\langle u^m_\textbf{k}|\epsilon|u^n_{\textbf{k}}\rangle=\delta_{mn}$ ($\delta_{mn}$ for Kronecker delta), $\Delta_x={2\pi}/{aN_x}\hat{x}$ and $N_x$ is the number of unit cells along $x$ direction, and $m, n\in 1, 2...N_{\rm{occ}}$ denote band
indices below the energy gap of interest~\cite{he2020, benalcazar2017electric}. We note that the inner product in Eq.~\eqref{U41} is defined the same as in Eq.~\eqref{Fmn},and the eigenmode normalization with $\epsilon(r)$ is used consistently throughout all our calculations, as outlined in Eqs. (A7) and (A13). Accordingly, the Wilson loop is defined as
\begin{eqnarray}
W_{x,\textbf{k}}\equiv W_{x,\textbf{k}+2\pi\hat{x}\leftarrow \textbf{k}}=F_{x,\textbf{k}+2\pi\hat{x}-\vartriangle_x}...F_{x,\textbf{k}+\vartriangle_x}F_{x,\textbf{k}}.\nonumber\\
\label{14}
\end{eqnarray}
The Wannier center $v^j_x(k_y)$ can be obtained by solving the eigenvalues of $W_{x,\textbf{k}}$, which corresponds to the average positions of the wave functions relative to the center of the unit cell: 
\begin{eqnarray}
W_{x,\textbf{k}}|v^j_{x,\textbf{k}}\rangle=e^{2\pi iv^j_x(k_y)}|v^j_{x,\textbf{k}}\rangle. 
\label{vx}
\end{eqnarray}
The eigenstates $|v^j_{x,\textbf{k}}\rangle$ have components $[v^j_{x,\textbf{k}}]^n$, where $n=1,2,...,N_{\rm{occ}}$. The polarization $p_x(k_y)$ along the $x$ direction can be obtained by
summing over all the Wannier bands below the band gap as 
\begin{eqnarray}
p_x(k_y)=\sum^{N_{occ}}_{l=1}v^l_x(k_y)=-\frac{i}{2\pi}\ln \det W_{x,\textbf{k}}.\label{px}
\end{eqnarray}
One gets the total polarization of the system along the $x$ direction $P_x$ by integrating over the
momentum $k_y$:
\begin{eqnarray}
P_x=\frac{1}{2\pi}\int^{2\pi}_0dk_yp_x(k_y).
\end{eqnarray}
Similarly, the total polarization along $y$ direction $P_y$ can also be calculated by the same analysis.

When the total polarization vanishes $P_i=0$ and the Wannier band has a band gap, the quadrupole moment can be defined. To characterize the quadrupole phase, we need to construct a new basis,
\begin{eqnarray}
|w^j_{x,\textbf{k}}\rangle=\sum_{n=1}^{N_{\rm occ}}[v^j_{x,\mathbf{k}}]^n|u^n_{\textbf{k}}\rangle,
\end{eqnarray}

The nested Wilson line along the $y$ direction for a Wannier sector $v^j_x(k_y)$ could be defined as
\begin{eqnarray}
\tilde{F}^{v_x}_{y,\textbf{k}}=\langle w^j_{x,\textbf{k}+\vartriangle_y}|\epsilon(\mathbf{r})|w^{j'}_{x,\textbf{k}}\rangle,
\end{eqnarray}
where $j,j' \in 1...N_w$ are all the Wannier bands within the Wannier sector $v_x$ in Eq.~\eqref{vx}. So the nested Wilson loop reads
\begin{eqnarray}
\tilde{W}^{v_x}_{y,\textbf{k}}=\tilde{F}_{y,\textbf{k}+2\pi\hat{y}-\vartriangle_y}...\tilde{F}_{y,\textbf{k}+\vartriangle_y}\tilde{F}_{y,\textbf{k}}.
\end{eqnarray}
One gets the polarization of the Wannier band by solving the nested Wilson loop eigenvalues in 
\begin{eqnarray}
\tilde{W}^{v_x}_{y,\textbf{k}}|v^{v_{x,j}}_{y,\textbf{k}}\rangle=e^{2\pi ip_y^{v_{x,j}}(k_x)}|v^{v_{x,j}}_{y,\textbf{k}}\rangle.
\end{eqnarray}
And the nested Wannier band polarization is
\begin{eqnarray}
p^{v_x}_{y}(k_x)=\sum^{N_{w}}_{j=1}p^{v_{x,j}}_y(k_x)=-\frac{i}{2\pi}\ln \det \tilde{W}^{v_{x}}_{y,\textbf{k}}. \label{pyvx}
\end{eqnarray}
Then the polarization $P^{v^j_x}_{y}$ of $j$-th Wannier band could be obtained by integrating over $k_x$, 
\begin{eqnarray}
P^{v^j_x}_{y}=\frac{1}{2\pi}\int^{2\pi}_0dk_xp^{v^j_x}_{y}(k_x).
\end{eqnarray}
Similarly, we can also get the polarization $P^{v^j_y}_{x}$ of the Wannier sector $v^j_y(k_x)$. 

The quadrupole moment can be defined by the polarization of the Wannier sector as
\begin{eqnarray}
q_{xy}=\sum^{N_{\rm{occ}}}_{j=1}P^{v^j_x}_yP^{v^j_y}_x. \label{quadrupole1}
\end{eqnarray}
Additional note that the quadrupole moment of the band gap could also be calculated by \cite{liu2019}
\begin{eqnarray}
q_{xy}=\sum^{N_{occ}}_{n}p^n_xp^n_y\quad \rm{mod} \quad 1,\label{quadrupole2}
\end{eqnarray}
where the $p^n_x$ and $p^n_y$ are not the total polarization $P_i$, but the polarization along two directions of $n-$th band. In this work, we use the two methods to define the quadrupole moment of the Lieb PhC.  

\begin{figure*}[hbtp!] 
  \includegraphics[width=0.7\textwidth]{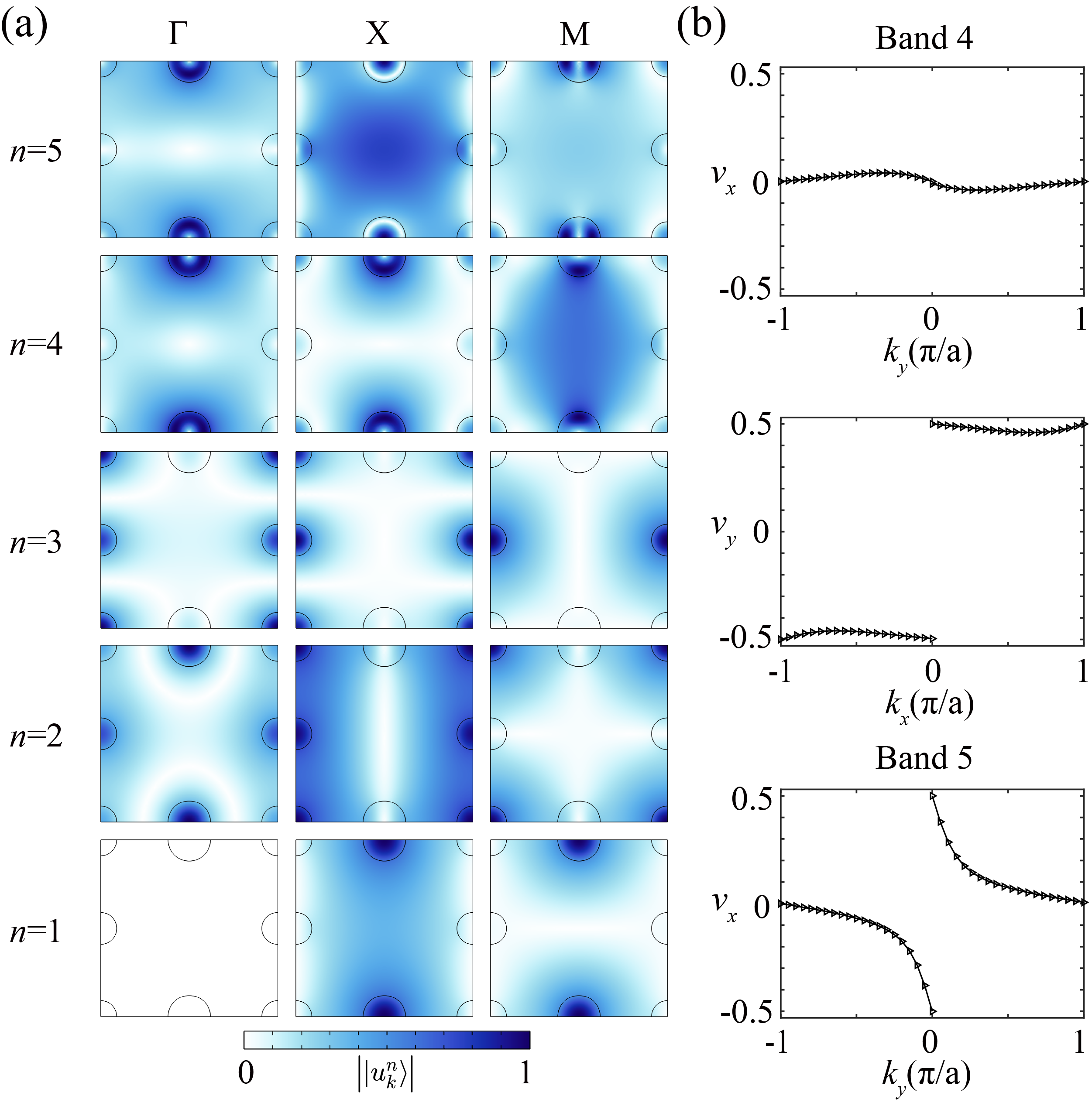}
\caption{\label{fig:figA2}(a) Bloch function magnitude $\big\vert\vert u_n(x, y)\rangle\big\vert(n=1\sim 5)$ at three high symmetry points: $\Gamma$, X and M respectively. (b) The Wannier center $v_x(k_y)$ and $v_y(k_x)$ of bands 4 and 5. Also cf.~the frequency diagram for type \uppercase\expandafter{\romannumeral1} in Fig.~\ref{fig:fig3}.}
\end{figure*}

To enhance reproducibility, Bloch functions $\vert u_n(\mathbf{k})\rangle$ are plotted for the lowest five bands at three reciprocal points: $\Gamma$, X and M respectively in Fig.~\ref{fig:figA2}(a). Wannier centres $v_x(k_y), v_y(k_x)$ for bands 4 and 5 are also accompanied in Fig.~\ref{fig:figA2}(b). Explicitly, they all carry $M_x, M_y$ symmetry, and the Bloch function for $n=2$ at M point has the $M_{xy}$ symmetry. \\

\section{Quantized dipole, quadrupole moments with reflection and rotation symmetries}\label{symmetry}

For a system with reflection symmetry $M_x$, the Hamiltonian of the system satisfies
\begin{eqnarray}
{M}_xh_{\textbf{k}}{M}^{\dagger}_x=h_{M_x\textbf{k}},
\end{eqnarray}
where ${M}_x\textbf{k}={M}_x(k_x,k_y)=(-k_x,k_y)$. Under the ${M}_x$ reflection symmetry, the wave function satisfies
\begin{eqnarray}
{M}_x|u^n_{\textbf{k}}\rangle &=&|u^m_{M_x\textbf{k}}\rangle \langle u^m_{M_x\textbf{k}}\vert{M}_x|u^n_{\textbf{k}}\rangle \nonumber\\
&=&|u^m_{M_x\textbf{k}}\rangle B^{mn}_{M_x,\textbf{k}},
\end{eqnarray}
where $B^{mn}_{M_x,\textbf{k}}$ is a sewing matrix connecting $\textbf{k}$ and $M_x\textbf{k}$. We select a short section of Wilson loop line element $F^{pq}_{x,\textbf{k}}$, which fulfills
\begin{eqnarray}
B^{mp}_{M_x,\textbf{k}}F^{pq}_{x,\textbf{k}}{B^{qn}_{M_x,\textbf{k}}}^{\dagger}&=&B^{mp}_{M_x,\textbf{k}}\langle u^p_{k+\delta k_x}|u^q_{\textbf{k}}\rangle{B^{qn}_{M_x,\textbf{k}}}^{\dagger}\nonumber\\
&=&\langle u^m_{M_x\textbf{k}}|M_x|u^p_{\textbf{k}}\rangle\langle u^p_{k+\delta k_x}|M_x^{-1}|u^n_{M_x\textbf{k}}\rangle,\nonumber\\
\label{a}
\end{eqnarray}
where 
\begin{eqnarray}
{M}_x|u^p_{\textbf{k}}\rangle&=&|u^r_{M_x\textbf{k}}\rangle B^{rp}_{M_x,\textbf{k}},\nonumber\\
\langle u^p_{\textbf{k}+\delta k_x}|M_x^{-1}&=&B^{sp*}_{M_x,\textbf{k}+\delta k_x}\langle u^s_{M_x,\textbf{k}+\delta k_x}|.\label{b}
\end{eqnarray}
Taking Eq.~\eqref{b} into Eq.~\eqref{a}, one gets
\begin{eqnarray}
B^{mp}_{M_x,\textbf{k}}F^{p,q}_{x,\textbf{k}}{B^{qn}_{M_x,\textbf{k}}}^{\dagger}&={F^{mn}_{x,M_x\textbf{k}}}^{\dagger}=F^{mn}_{-x,M_x\textbf{k}}.\label{c}
\end{eqnarray}
We can further extend Eq.~\eqref{c} to the entire Wilson loop, as

\begin{eqnarray}
B^{mp}_{M_x,\textbf{k}}W^{p,q}_{x,\textbf{k}}{B^{qn}_{M_x,\textbf{k}}}^{\dagger}&={W^{mn}_{x,M_x\textbf{k}}}^{\dagger}=W^{mn}_{-x,M_x\textbf{k}}.\label{d}
\end{eqnarray}
From Eq.~\eqref{d}, the polarization $p_x$ for a lattice under $M_x$ reflection symmetry satisfies
\begin{eqnarray}
  p_x(k_y)\overset{M_x}{=}-p_x(k_y)\mod 1,
\end{eqnarray}
i.e.,
\begin{eqnarray}
p_x(k_y)\overset{M_x}{=}0\quad\text{or}\quad1/2.\label{e}
\end{eqnarray}
Also for $p_y(k_x)$, 
\begin{eqnarray}
p_y(k_x)&&\overset{M_x}{=} p_y(-k_x)\mod 1. \label{pykx}
\end{eqnarray}

Similarly, the polarization $p_x, p_y$ for a lattice with $M_y$ reflection satisfy
\begin{eqnarray}
p_x(k_y)\overset{M_y}{=}p_x(-k_y) &\mod 1,\\
p_y(k_x)\overset{M_y}{=}-p_y(k_x) &\mod 1.\label{f}
\end{eqnarray}
Accordingly the total polarization $p_x$ and $p_y$ with both reflection symmetries $M_x$ and $M_y$ are quantized, i.e.,
\begin{eqnarray}
p_x(k_y)\overset{M_x}{=}0 \quad\text{or}\quad1/2,\\
p_y(k_x)\overset{M_y}{=}0 \quad\text{or}\quad1/2.
\end{eqnarray}

The polarization of Wannier sectors is also quantized with these symmetries~\cite{benalcazar2017electric}, i.e.,
\begin{eqnarray}
p_y^{v_x}\overset{M_x}{=}p_y^{-v_x} \mod 1,\\
p_y^{v_x}\overset{M_y}{=}-p_y^{v_x} \mod 1.
\end{eqnarray}

Moreover, for a system with $C_4$ rotation symmetry, the polarizations satisfies 
\begin{eqnarray}
p_x(k_y)\overset{C_4}{=}p_y(k_x=-k_y)\mod 1,\\
p_y(k_x)\overset{C_4}{=}-p_x(k_y=k_x)\mod 1,
\end{eqnarray}
then one gets
\begin{eqnarray}
p_x(k_y)\overset{C_4}{=}p_y(k_x)\overset{C_4}{=}0 \quad\text{or}\quad1/2. \label{C4}
\end{eqnarray}
When we change the distance between sublattices A, B and C simultaneously, i.e., $\delta d_A=\delta d_C$, the $M_x$ and $M_y$ reflection symmetries are broken, but maintain reflection symmetry $M'$ along the diagonal $y=x$. Under the $M'$ reflection symmetry, following two directions $x, y$ of Wilson loop, the polarization satisfies
\begin{eqnarray}
p_x(k_y)\overset{M'}{=}p_y(k_x) \mod 1,
\end{eqnarray}
which shows the polarization along $k_x$ direction is the same as that along $k_y$ direction. Furthermore, there are other constraints to make it quantized, 
\begin{equation}
p_{x'}\overset{M'}{=}0\quad \text{or}\quad 1/2. \label{pxp} 
\end{equation}
The quantized result for $p_{x'}$ in Eq.~\eqref{pxp} can be obtained after rotating by $\pi/4$ the 2D coordinate $k_x, k_y$ to $k'_x, k_y'$ as shown in Fig.~\ref{fig:figA1}. Meanwhile it is also accessible by directly following Subsec. 1, Append. D in ~\cite{benalcazar2017electric}. 
The essential building block is the sewing matrix $B_{g, \mathbf{k}}^{mn}$, which is defined as Eq.~\eqref{a} therein
\begin{equation}
B_{g, \mathbf{k}}^{mn}=\langle u^m_{D_g \mathbf{k}} \vert g_{\mathbf{k}} \vert u^n_{\mathbf{k}} \rangle.  
\end{equation}
And it absorbs reciprocal lattice vector $\mathbf{G}$:   
\begin{equation}
B_{g, \mathbf{k}+\mathbf{G}}^{mn}=B_{g, \mathbf{k}}^{mn}
\end{equation}
by using $\vert u^n_{\mathbf{k}+\mathbf{G}} \rangle = V(-\mathbf{G}) \vert u^n_k\rangle$.  
What differs here is the term $\mathbf{G}$ added to the momentum $\mathbf{k}$: it traverses along $x'$ axis in Fig.~\ref{fig:figA1}. Numerically one follows the procedure in Append.~\ref{Method} to obtain Wannier centers similarly, in which a translation operation is used to facilitate it along the slanted $x'$ direction~\cite{jin2017infrared}. 

\section{Gap depth dependence for the edge and corner states} \label{air gap}

\begin{figure*}[hbtp!]
\includegraphics[width=0.8\textwidth]{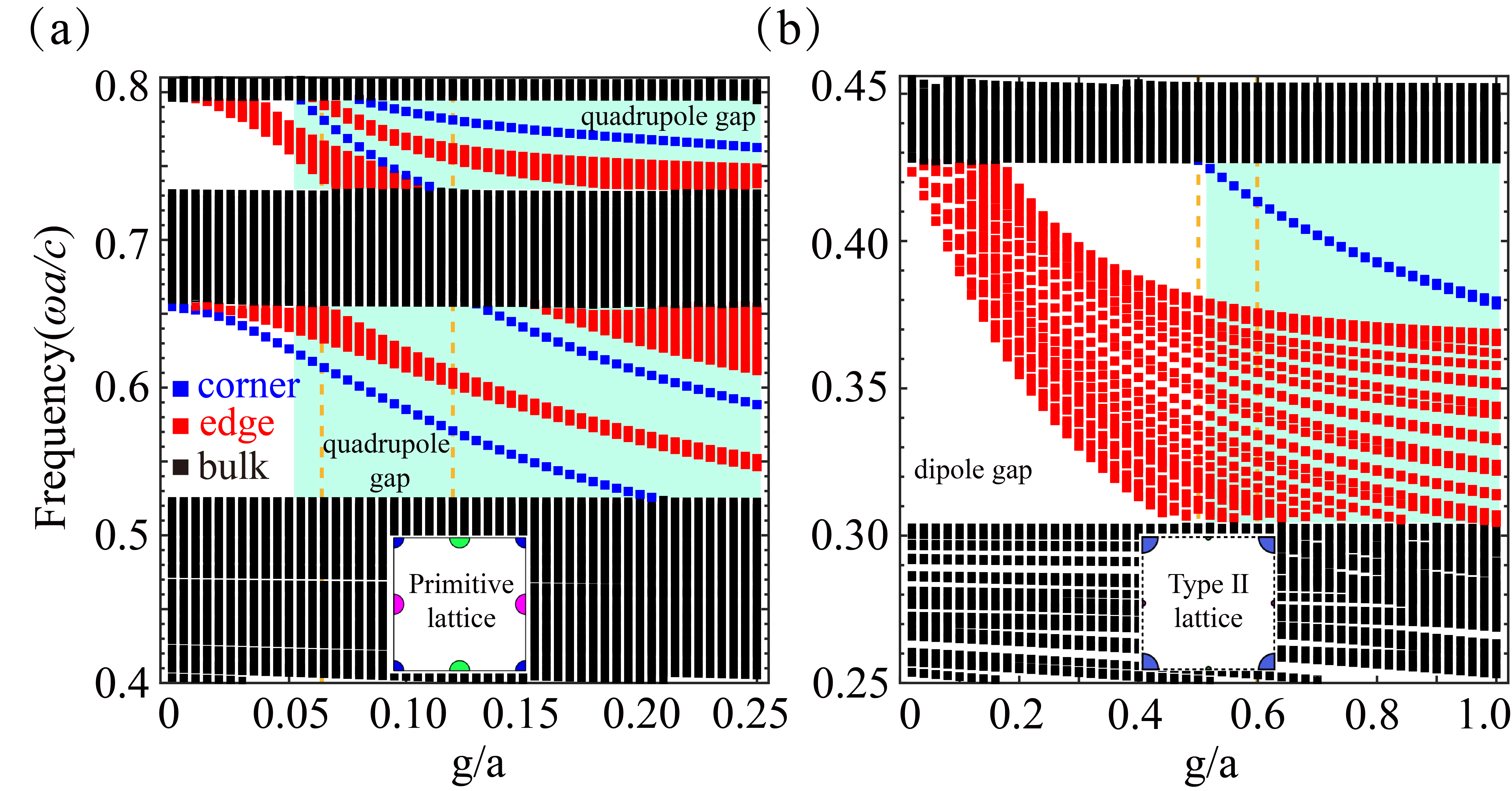}
\caption{\label{fig:figA3}(a) The frequency diagram for Fig.~\ref{fig:fig2}(d) as a function of $g$, along with the edge (in red) and corner states (in blue) in a quadrupole gap. (b) For Fig.~\ref{fig:fig4}(d) including the edge and corner states in a dipole gap. The insets represent the unit cell for each panel, and both for primitive and type \uppercase\expandafter{\romannumeral2} ($10\times10$ lattices). Note the parameter chosen for Figs.~\ref{fig:fig2} and~\ref{fig:fig4} is indicated by the dashed orange line in each panel.  
}
\end{figure*}

In Append.~\ref{air gap}, we demonstrate the numerical results for the frequency range of edge and corner states with respect to the bulk. The positions of edge and corner states caused by dipole and quadrupole moments can be changed by tuning the air gap width. Specifically, we change the width of air gap $g$ to control the boundary condition. As $g$ increases, the frequency of edge states and corner states gradually decreases or even immerses into the bulk band, as shown in Figs.~\ref{fig:figA3}(a) and~\ref{fig:figA3}(b).

Note that if the air gap width is set as $g=0$, the edge states and corner states caused by the dipole and quadrupole moments will be immersed in the bulk band, which is not legible in gap as shown in Fig.~\ref{fig:figA3} for two cases. Similarly, for the super cell consisting of $10\times 10$ unit cells as shown in Fig.~\ref{fig:fig3}(e)[data not shown], the air gap is also required to tweeze in-gap the edge and the corner states. 

As for the \emph{physical principle of} depth $g$ to regulate edge/corner states, the air gap serves as a tunable boundary potential that controls the coupling between the photonic crystal and the external environment (modeled as a PEC). A deep air gap (large $g$) increases the effective potential barrier at the boundary, pushing edge/corner states deeper into the bulk band gap and thus enhancing their spatial localization. As $g$ varies, so does the boundary condition exerted in calculation, and the eigenfrequencies of edge/corner states are shifted within the gap. The optimal ranges of air gap depths are exemplified in Fig.~\ref{fig:figA3} for both cases.

\begin{acknowledgments} 
Z.-K. X., and Y. L. thank Liu Zheng-Rong, Chen Rui, Wei Xian-Hao and Lan Zhihao for useful discussion. 
The work was supported by National Natural Science Foundation of China (Grants Nos. 11804087, 12074107, 62571212, 12374415, U25D8012), Natural Science Foundation of Hubei Province of China (Grants Nos. 2024AFA038, 2022CFB553, 2022CFA012, 2023AFB917), Wuhan City Key R\&D Program (Grant No. 2025050602030069), and Program of Outstanding Young and Middle-aged Scientific and Technological Innovation Team of Colleges and Universities in Hubei Province of China (Grant No. T2020001).
X. F. is also supported by Chutian Scholars Program in Hubei Province.

\end{acknowledgments}


\newpage
\bibliography{reference}
\end{document}